\documentclass[apj,twocolumn]{aastex63}
\usepackage[mathscr]{eucal}
\usepackage{amsmath}
\usepackage{systeme}

\shorttitle{Tracing cluster properties along a common evolutionary track spanning 9 Gyr}
\shortauthors{F. Ruppin \emph{et al.}}

\def\araa{ARA\&A}%
\def\apj{ApJ}%
\def\apjl{ApJ}%
\def\apjs{ApJS}%
\def\aap{A\&A}%
\def\mnras{MNRAS}

\def\nat{Nature}
\def\jcap{JCAP}

\def\xmm{XMM-{\it Newton}}
\def\planck{{\it Planck}} 
\def\chandra{{\it Chandra}}

\newcommand{\MIT}{Kavli Institute for Astrophysics and Space Research, Massachusetts Institute of Technology, 77 Massachusetts Avenue, Cambridge, MA 02139}
\newcommand{\ANL}{High Energy Physics Division, Argonne National Laboratory, 9700 South Cass Avenue, Lemont, IL 60439, USA}
\newcommand{\KChic}{Kavli Institute for Cosmological Physics, University of Chicago, 5640 South Ellis Avenue, Chicago, IL 60637, USA}
\newcommand{\UChic}{Department of Astronomy and Astrophysics, University of Chicago, 5640 South Ellis Avenue, Chicago, IL 60637, USA}
\newcommand{\UMiss}{Department of Physics and Astronomy, University of Missouri, 5110 Rockhill Road, Kansas City, MO 64110, USA}
\newcommand{\KStan}{Kavli Institute for Particle Astrophysics \& Cosmology, P. O. Box 2450, Stanford University, Stanford, CA 94305, USA}
\newcommand{\UStan}{Department of Physics, Stanford University, 382 Via Pueblo Mall, Stanford, CA 94305, USA}
\newcommand{\SLAC}{SLAC National Accelerator Laboratory, 2575 Sand Hill Road, Menlo Park, CA 94025, USA}
\newcommand{\FLab}{Fermi National Accelerator Laboratory, P. O. Box 500, Batavia, IL 60510, USA}

\begin{document}

\title{Stability of Cool Cores During Galaxy Cluster Growth: A Joint {\chandra}/SPT Analysis of 67 Galaxy Clusters Along a Common Evolutionary Track Spanning 9 Gyr}

\author{F.~Ruppin}\affiliation{\MIT}
\author{M. McDonald}\affiliation{\MIT}
\author{L. E. Bleem}\affiliation{\ANL}\affiliation{\KChic}
\author{S.~W. Allen}\affiliation{\KStan}\affiliation{\UStan}\affiliation{\SLAC}
\author{B.~A. Benson}\affiliation{\FLab}\affiliation{\KChic}\affiliation{\UChic}
\author{M. Calzadilla}\affiliation{\MIT}
\author{G. Khullar}\affiliation{\UChic}
\author{B. Floyd}\affiliation{\UMiss}

\correspondingauthor{F. Ruppin}
\email{ruppin@mit.edu}

\begin{abstract}

We present the results of a joint analysis of \chandra\ X-ray and South Pole Telescope (SPT) SZ observations targeting the first sample of galaxy clusters at $0.3 < z < 1.3$, selected to be the progenitors of well-studied nearby clusters based on their expected accretion rate. We develop a new procedure in order to tackle the analysis challenge that is estimating the intracluster medium (ICM) properties of low-mass and high-redshift clusters with ${\sim}150$ X-ray counts. One of the dominant sources of uncertainty on the ICM density profile estimated with a standard X-ray analysis with such shallow X-ray data is due to the systematic uncertainty associated with the ICM temperature obtained through the analysis of the background-dominated X-ray spectrum. We show that we can decrease the uncertainty on the density profile by a factor ${\sim}5$ with a joint deprojection of the X-ray surface brightness profile measured by \chandra\ and the SZ integrated Compton parameter available in the SPT cluster catalog. We apply this technique to the whole sample of 67 clusters in order to track the evolution of the ICM core density during cluster growth. We confirm that the evolution of the gas density profile is well modeled by the combination of a fixed core and a self-similarly evolving non-cool core profile. We show that the fraction of cool-cores in this sample is remarkably stable with redshift although clusters have gained a factor ${\sim}4$ in total mass over the past ${\sim}9$~Gyr. This new sample combined with our new X-ray/SZ analysis procedure and extensive multi-wavelength data will allow us to address fundamental shortcomings in our current understanding of cluster formation and evolution at $z > 1$.

\end{abstract}


\section{Introduction}\label{sec:intro}

\begin{figure*}[t]
\centering
\includegraphics[height=9cm]{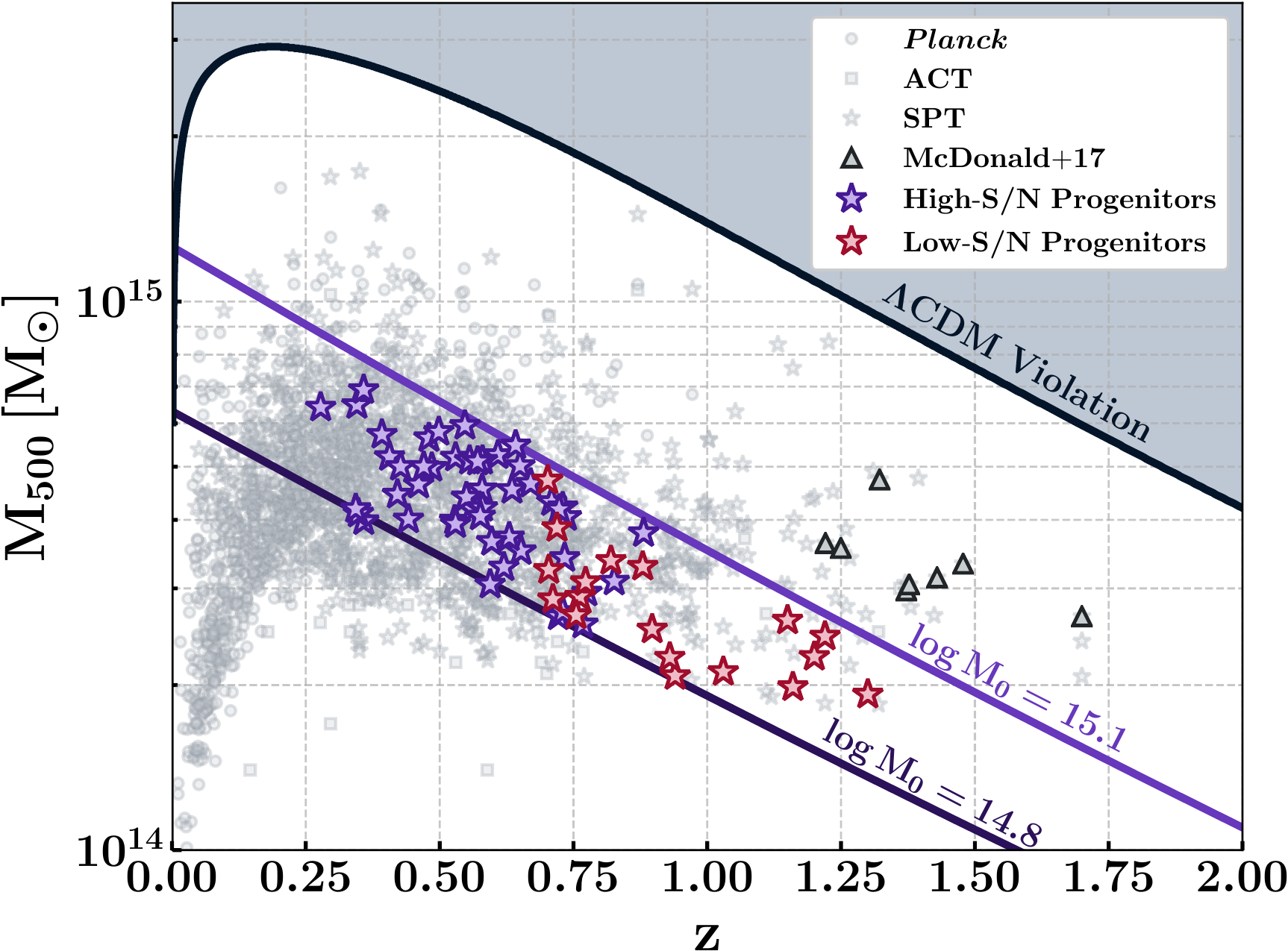}
\caption{{\footnotesize Location of each cluster of the considered sample in the mass-redshift plane. We sub-divided the sample into a high-S/N sample (purple stars) and a low-S/N one (red stars) based on the number of available \chandra\ counts. Diagonal lines show the predicted growth for halos of different final masses \citep{fak10}. The 8 high redshift SPT clusters analyzed by \cite{mcd17} are shown with grey triangles for comparison along with the \planck, ACT, and SPT clusters (light grey symbols). We also indicated the 90\% confidence level upper limit on the halo mass at each redshift given the considered cosmological model (black line).}}
\label{fig:m_z_plane}
\end{figure*}

Galaxy clusters are the end result of a hierarchical process starting from matter density peaks at the end of inflation that first grew through the smooth accretion of surrounding material \citep[\emph{e.g.}][]{pre74,coo02,mea15}. Merger events with smaller halos then contributed to both the galaxy cluster growth and the heating of their baryonic matter content called the intracluster medium (ICM) up to few keV \citep[\emph{e.g.}][]{sar02,mar07,bou13}. Studying the evolution of the ICM thermodynamic properties with cluster redshift and mass is essential to unveiling the multi-phase and multi-scale physical mechanisms at play during their growth \citep[\emph{e.g.}][]{voi08,voi15,mcn16,tum19,gas20}. Such understanding is key to use galaxy clusters as tracers of the history of large scale structure formation \citep[\emph{e.g.}][]{voi05,pla15,val20} and as probes of the underlying cosmology \citep[\emph{e.g.}][]{all11,has13,boc15,pla16,hil18,boc19}.\\
\indent Unveiling the properties and evolution of the ICM and the active galactic nucleus (AGN)-star formation-halo connection in early-forming systems all the way back to $z{\sim}3$ will be among the primary science goals of both \emph{Athena} \citep{bar20} and \chandra\ successor missions such as \emph{Lynx} \citep{lyn18} or the Advanced X-ray Imaging Satellite \citep{mus19}. While many of the most exciting questions about the initial formation of galaxy clusters must wait for these next-generation X-ray missions, the current X-ray observatories can lay an important foundation now by studying clusters in the $1 < z < 2$ range, where to date only a dozen of the most massive systems have been observed.\\
\indent Until recently, studies of distant galaxy clusters were limited to a small number of extreme systems, discovered serendipitously in deep X-ray observations \citep[\emph{e.g.}][]{sch04,kol06,fin10}. However, the successes of Sunyaev-Zel'dovich (SZ) surveys \citep{pla16b,has13,hil18,ble15,hua19,ble19,hil20} have rapidly altered the landscape of galaxy cluster astrophysics and cosmology. In particular, the South Pole Telescope \citep[SPT;][]{car11} has surveyed 5000 deg$^2$ of the southern sky over the past 10 years, leading to the discovery of 1066 galaxy clusters, including 72 at $z>1$. The combination of SPT selection, which is redshift independent and only limited by the survey sensitivity, with relatively shallow \chandra\ follow-up has proven an extremely efficient way of studying the growth and evolution of the most massive clusters \citep[\emph{e.g.}][]{mcd13,mcd14,mcd16,mcd17}.\\
\indent In particular, \cite{mcd17} studied 8 massive SPT clusters at $z > 1$ (see Fig.~\ref{fig:m_z_plane}) based on 892~ks of \chandra\ observations in order to measure ${\sim}1500$ counts for each cluster. One of the main results of this work is that cluster cool cores have had fixed properties in the past 10~Gyr while the bulk of their halo kept growing self-similarly around them. While interesting in their own right, these systems may, however, experience a different evolution than the clusters we are most familiar with at $z{\sim}0$ (\emph{e.g.}, Abell\, 2390 \citep{all01}, Perseus \citep{fab11}, Zwicky\, 3146 \citep{rom20}, etc.), which were considerably less massive at $z>1$. Based on the Millennium-II simulations, \cite{fak10} found that a typical cluster will increase in mass by a factor of $\sim$4 over the past $\sim$9 Gyr (see Fig.~\ref{fig:m_z_plane}). So, to study the evolution of a sample of clusters at $z=0$ with M$_{500}{\sim}8\times 10^{14}$~M$_{\odot}$, we should be comparing to systems at $z{\sim}1.4$ with M$_{500}{\sim}2 \times 10^{14}$~M$_{\odot}$.\\
\indent At such high redshifts, a radial distance of 50~kpc corresponds to a projected angle of 6~arcsec which is about the size of \xmm\ point spread function \citep{lum12}. Furthermore, the small angular size of these high-redshift and low-mass clusters makes point source contamination of the extended emission a bigger issue than it is at low redshift. \chandra\ is therefore the most appropriate observatory to consider in order to characterize the core properties of these systems. Unfortunately, the typical \chandra\ X-ray count rate of such clusters is ${\sim}2\times 10^{-3}$ count/s in the 0.7-2~keV band. Measuring the same number of counts as in \cite{mcd17} would therefore require exposures of about 1~Ms per cluster. The high cost of high-$z$ cluster observations can however be driven by the need to measure a mean ICM temperature from X-ray spectroscopy with relative uncertainties of the order of ${\sim}10\%$ in order to measure an accurate ICM density profile. Joint X-ray/SZ analyses have recently proven their high efficiency in characterizing the ICM properties of massive $z>1$ clusters with relatively shallow X-ray observations \citep[\emph{e.g.}][]{ada17,ghi18,rup20,cas20,ker20}.\\
\indent In this paper, we build upon this past experience and analyze jointly low signal-to-noise (S/N) \chandra\ data and the SPT SZ signal measured in a sample of 67 clusters selected to be the progenitors of more common and well-studied systems at $z{\sim}0$ (\emph{i.e.} M$_{500}(z=0) \sim 8 \times 10^{14}$~M$_{\odot}$). Our goal is to demonstrate that the combination of ${\sim}150$ \chandra\ counts and the integrated SZ signal from SPT can provide sufficient constraining power to estimate the ICM density profile of $z>1$ low-mass clusters. We describe the selection procedure of the 67 SPT clusters in \textsection \ref{sec:spt_prog_samp} as well as the \chandra\ observations. In \textsection \ref{sec:xray_sz}, we present the details of the analysis methodology considered in this work based on \chandra\ X-ray and SPT SZ data. We characterize the performance of the joint X-ray/SZ analysis procedure and emphasize the information gain with respect to a standard X-ray analysis in \textsection \ref{sec:performance}. We describe the results on the ICM properties obtained for the whole sample in \textsection \ref{sec:results}. In \textsection \ref{sec:discussion}, we discuss these results in the context of cluster evolution and highlight the legacy value of this particular sample. We give a summary of our work in \textsection \ref{sec:conclu}.  In this paper, we assume a flat $\mathrm{\Lambda}$CDM cosmology with $\Omega_m = 0.3$, $\Omega_{\Lambda} = 0.7$, $H_0 = 70~\mathrm{km\,s^{-1}\,Mpc^{-1}}$, and define the radius R$_{500}$ and mass M$_{500}$ in terms of the critical density $\rho_c(z)$ at the cluster redshift $z$ as: $\mathrm{M}_{500} \equiv \frac{4\pi}{3}500\rho_c(z)\mathrm{R}_{500}^3$.


\section{SPT progenitor sample}\label{sec:spt_prog_samp}

Studying mass-selected samples of galaxy clusters at $z>1$ that are the progenitors of typical halos at $z{\sim}0 $ is crucial to our understanding of cluster evolution. The \chandra\ follow-up of ${\sim}100$ clusters \citep{mcd13} from the first generation SPT cluster catalog yielded tremendous scientific returns, including, among others, the discovery and characterization of the Phoenix cluster \citep{wil11,mcd12,mcd13b,mcd15,mcd19}, the evolution of radio-mode feedback \citep{hla15}, and the evolution of the ICM metallicity \citep{mcd16,man17,man20}. However, such mass-selected samples of clusters contain systems that follow different evolutionary tracks. On average, the massive clusters that have been observed so far at $z>1$ do not evolve into the well-known intermediate mass clusters at $z{\sim}0$. Unlike previous generations of high-$z$ cluster surveys \citep[\emph{e.g.}][]{mcd17,san17}, our goal is no longer to target rare, extreme systems at $z>1$, but instead to characterize the progenitors of present-day clusters like Perseus \citep{fab11} and Abell\, 2390 \citep{all01}. To this end, we define a new sample of clusters in order to characterize how their properties evolve during their growth in the past 9~Gyr. This section first describes the selection procedure that we followed to obtain this sample. We then present the \chandra\ observations realized for each cluster in the sample and study their quality in terms of S/N.

\begin{figure*}[t]
\centering
\includegraphics[height=6.8cm]{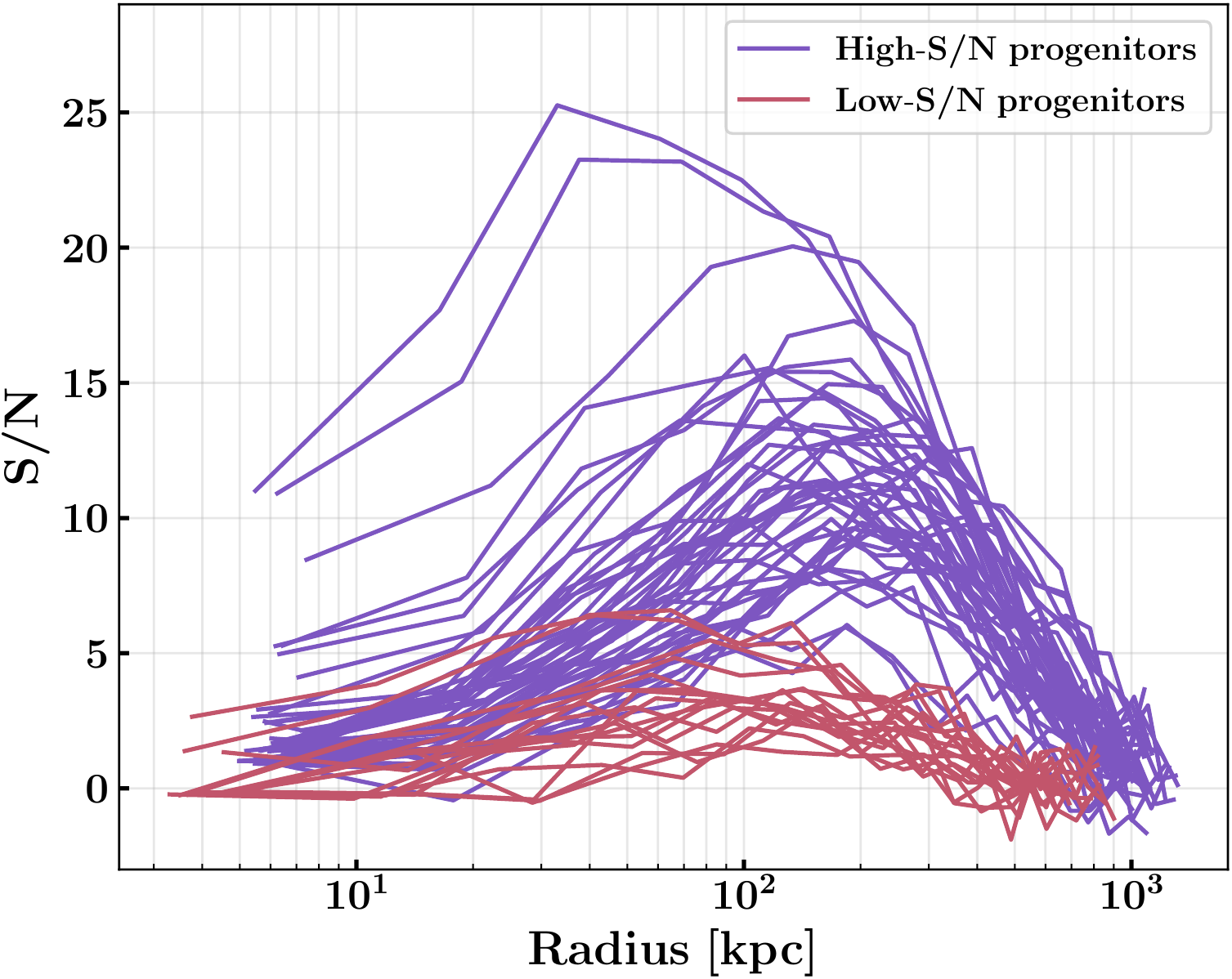}
\hspace{0.5cm}
\includegraphics[height=6.8cm]{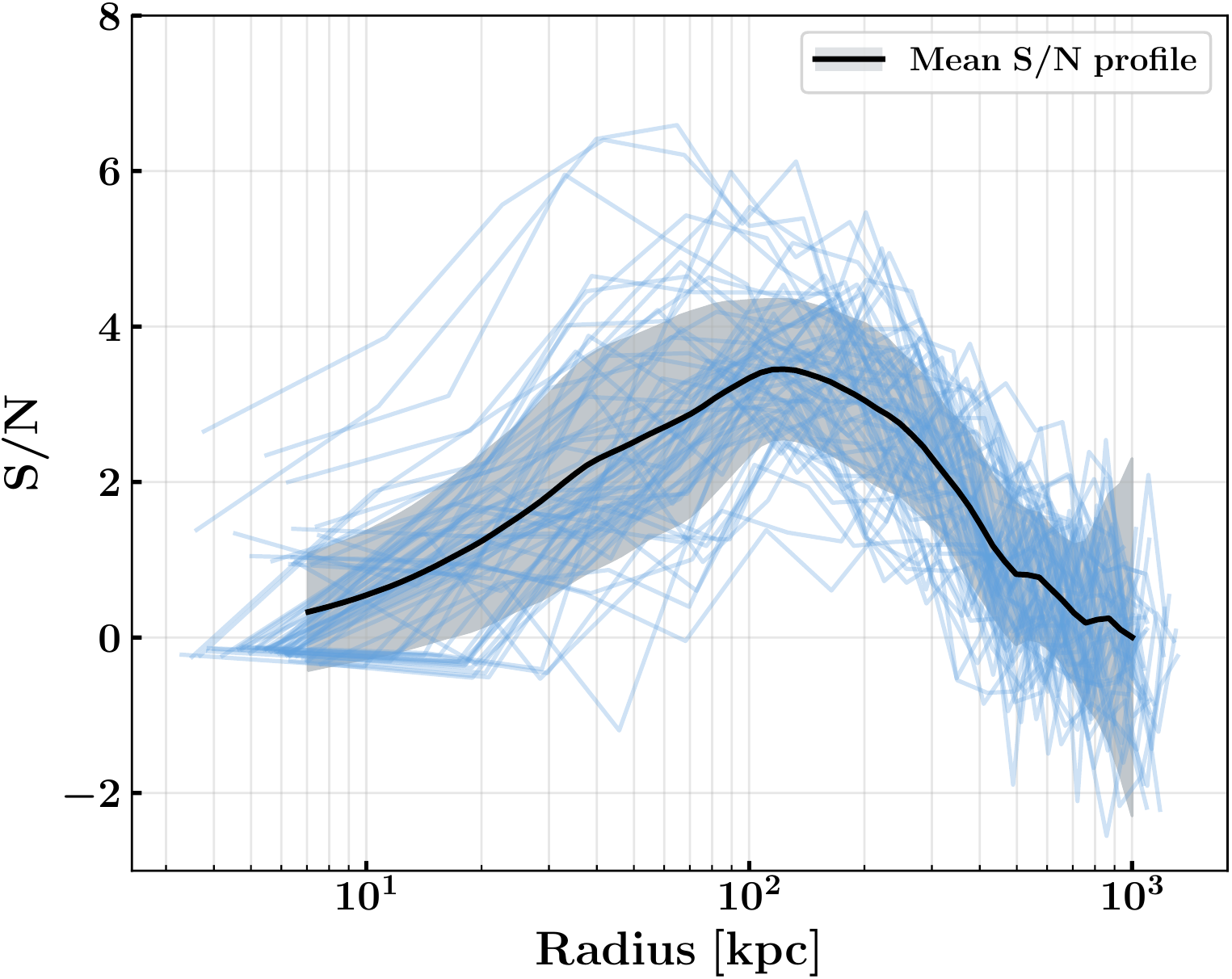}
\caption{{\footnotesize \textbf{Left:} Signal-to-Noise (S/N) profiles estimated in the \chandra\ 0.7-2 keV band from the event files obtained for each cluster in the considered sample, see Eq. (\ref{eq:snr_prof}). S/N profiles of clusters classified in the high-S/N sub-sample are shown in purple and those from the low-S/N one are shown in red. \textbf{Right:} S/N profiles of all 67 clusters after applying the re-scaling procedure described in \textsection \ref{subsec:rescale} to all clusters in the high-S/N sub-sample. The mean and scatter of the distribution of profiles are shown with the black line and grey area, respectively.}}
\label{fig:snr_prof}
\end{figure*}

\subsection{Selection procedure}\label{subsec:selection}

Although smooth accretion of matter largely participates in the growth of initial density fluctuations, the dominant channel for mass growth of galaxy clusters is given by the merging events with sub-halos \citep[\emph{e.g.}][]{gen10,ich15,sch19,val20}. By constructing the merger trees of dark matter haloes using a joint dataset from the Millennium \citep{spr05} and Millennium-II \citep{boy09} simulations, \cite{fak10} established an analytic formula for the mean mass growth rate of haloes as a function of redshift and mass in a wide range of descendant halo mass ($10^{10} < M_0 < 10^{15}$~M$_{\odot}$):
\begin{equation}
\begin{split}
\left\langle \frac{dM}{dt} \right\rangle =\, & 46.1~\mathrm{M_{\odot}yr^{-1}} \left( \frac{M}{10^{12}~\mathrm{M_{\odot}}} \right)^{1.1} \\
& \times (1 + 1.11z)\sqrt{\Omega_m (1+z)^3 + \Omega_{\Lambda}}
\end{split}
\label{eq:mass_growth}
\end{equation}
where $M$ is the halo mass at time $t$ and redshift $z$.\\
\indent We have used Eq. (\ref{eq:mass_growth}) in order to define an interval of evolutionary tracks that result in galaxy clusters at $z=0$ with masses ranging from M$_{500} = 6.3 \times 10^{14}$~M$_{\odot}$ to M$_{500} = 1.3 \times 10^{15}$~M$_{\odot}$ in order to characterize the evolution of typical clusters such as Perseus and Abell\, 2390. They are presented as light and dark purple lines in Fig.~\ref{fig:m_z_plane}. We notice that the 8 clusters studied in \cite{mcd17} at $z>1$ (grey triangles) correspond to systems that eventually evolve on average into clusters with M$_{500} > 1.3 \times 10^{15}$~M$_{\odot}$ at $z=0$. Such clusters have not been found in our local ($z < 0.1$) universe due to the limited comoving volume available (see black line in Fig.~\ref{fig:m_z_plane}) and therefore correspond to extreme cases.\\
\indent In this study, we select a sample of clusters that are the progenitors of existing well-known clusters at $z{\sim}0$ by considering halos that fall between the two boundaries defined by considering the range of descendant halo masses $[6.3 - 13.0]\times 10^{14}$~M$_{\odot}$ in Eq. (\ref{eq:mass_growth}). We use the SPT cluster catalogs as they currently contain the highest number of SZ-selected clusters at $z > 1$ with M$_{500} \sim 2 \times 10^{14}$~M$_{\odot}$. In particular, with the completion of a deep 100 deg$^2$ survey by the second-generation SPT camera, SPTpol, the SPT collaboration established the only SZ catalog with available clusters passing our selection criteria at $z>1.1$ given the considered range of descendant halo mass. Therefore, we consider both the SPT-SZ \citep{ble15} and SPTpol 100d \citep{hua19} catalogs to define our cluster sample. Throughout this study, we will only consider the masses M$_{500}$ and redshifts $z$ given in these two catalogs to define the sample and to characterize the ICM properties of each cluster.\\
\indent Among the 83 SPT-SZ clusters studied by \cite{mcd13} with \chandra, 49 systems satisfy our selection cuts. These clusters span a redshift range $0.3 \lesssim z \lesssim 0.8$ and are represented with purple stars in Fig.~\ref{fig:m_z_plane}. Dedicated \chandra\ observations have been realized in order to follow-up 18 clusters from the SPTpol 100d catalog (see \textsection \ref{subsec:x_ray_obs}) and increase the redshift range covered by this sample up to $z=1.3$. These clusters are shown with red stars in Fig.~\ref{fig:m_z_plane}.\\
\indent From Eq.~(\ref{eq:mass_growth}), we expect all 67 clusters in this sample to have similar mass growth rates on average and to eventually evolve statistically into the most well-studied clusters at $z{\sim}0$. Therefore, this sample provides a unique opportunity to characterize the evolution of the ICM thermodynamic properties during cluster growth and compare it with the outputs from hydrodynamic simulations that have access to the formation history of each halo in the simulated volume \citep[\emph{e.g.}][]{nel15,bar18,rup19}.\\
\indent Critically, the 67 clusters in this sample have a rich and scientifically enabling multiwavelength coverage, with nearly all having observations with DECam (grizY), \emph{Spitzer} (3.6, 4.5$\mu$m), \emph{Herschel} SPIRE (250, 300, 500$\mu$m) and ATCA (2.1 GHz). Therefore, the full sample of 67 progenitor-selected clusters represents a unique proving ground for the new multi-wavelength science investigations that will begin in the 2020's, with analogs to every data set that will become available (CMB-S4, Nancy Grace Roman Space Telescope, SKA, and Rubin Observatory) already in hand.

\begin{table*}
\begin{tabular}{cccccccc}
\hline
\hline
 & Number of Clusters & X-ray Program & $\langle z \rangle$ & $\langle S/N \rangle$ & Validation Sample \\
 \hline
high-S/N & 50 & XVP \citep{mcd13} & $0.57$ & 12 & Yes\\
 \hline
low-S/N & 17 & This work & $0.94$ & 3 & No\\
\hline
\end{tabular}
\caption{{\footnotesize General information concerning the high-S/N and low-S/N sub-samples considered in this work including: the number of clusters, the X-ray program, the average redshift and peak S/N, and whether the sub-sample is used to validate the joint X-ray/SZ procedure described in \textsection \ref{subsec:validation} or not.}}
\label{tab:summary}
\end{table*}

\subsection{Chandra X-ray observations}\label{subsec:x_ray_obs}

All 67 clusters considered in this work have been detected by SPT and have additionally been observed with \chandra. A sub-sample of 49 clusters from the SPT-SZ catalog have been observed through the Chandra X-ray Visionary Project (XVP; PI: B. Benson) described in \cite{mcd13}. This program has mostly been conducted during \chandra\ Cycles 12 and 13 with exposures typically sufficient to obtain ${\sim}1300$~counts per cluster in the 0.7-2~keV band (see Tab.~\ref{tab:high_sn} in Appendix~\ref{sec:app_A}). The remaining 18 clusters at $z>0.7$ selected from the SPTpol 100d catalog have been observed in the VFAINT data mode using the Advanced CCD Imaging Spectrometer (ACIS) I-chips on board \chandra\ in Cycles 18 to 20 with typical exposures ranging from 15 to 150~ks per cluster. These exposures have been chosen in order to reach a minimum of 110~counts per cluster and allowed us to obtain an average of 180 counts in the 0.7-2~keV band (see Tab.~\ref{tab:low_sn} in Appendix~\ref{sec:app_A}).

\subsection{Signal-to-noise profiles}\label{subsec:rescale}

We investigate the distribution of S/N profiles obtained from the \chandra\ observations as the two sub-samples of SPT-SZ and SPTpol clusters have not been observed with the same depth. As our ultimate goal is to characterize the redshift evolution of the ICM properties of the 67 clusters in our sample, it is essential that we use X-ray event files with similar S/N to avoid evolution biases caused by the evolving depth of the observations with halo mass and redshift.\\
\indent Therefore, we extract the total number of counts $N_{tot}$ as well as the number of counts due to background $N_B$ in the 0.7-2~keV band in different annuli centered on the X-ray centroid (see \textsection \ref{sec:xray_sz}) for each cluster in the sample. The S/N profiles are then computed with the following equation:
\begin{equation}
S/N = \frac{N_S}{\sqrt{N_{tot} + \sigma_B^2 \times \mathrm{n_{pix}}}}
\label{eq:snr_prof}
\end{equation}
where $N_S = N_{tot} - N_B$ is the number of counts due to signal, $\sigma_B$ is the per pixel uncertainty on the background estimate, and $\mathrm{n_{pix}}$ is the number of pixels in the considered annulus. The profiles are shown in the left panel of Fig.~\ref{fig:snr_prof} as a function of physical radius given the redshift of each cluster. We clearly see the distinction between the SPT-SZ clusters observed with \chandra\ in the context of the XVP program (purple) and the SPTpol clusters (red). The average S/N peak for the SPT-SZ clusters is about 12 while the one for the SPTpol systems is around 3. These two sub-samples will therefore be dubbed the high-S/N and low-S/N sub-samples throughout this paper. We note that one of the clusters from the SPTpol 100d catalog, SPT-CLJ0000-5748, has actually been observed in the context of an independent program (PI: Hlavacek-Larrondo) with a total number of counts of 4731 (see Tab.~\ref{tab:high_sn} in Appendix~\ref{sec:app_A}). There are therefore 50 clusters (instead of 49) in the high-S/N sub-sample. The mean readshift of the clusters in the high-S/N sub-sample is $\langle z \rangle = 0.57$ while the one of the low-S/N sub-sample is $\langle z \rangle = 0.94$.\\
\indent We create new event files from the ones obtained with the XVP program in order to scale all S/N profiles to a similar value for our final study of the redshift evolution of the ICM properties. This is done by measuring $N_S$ for each cluster in the high-S/N sub-sample in a circular region of radius R$_{500}$\footnote{Throughout this work, we use the SPT estimate of M$_{500}$ in \cite{ble15} and \cite{hua19} to compute R$_{500}$.} centered on the X-ray centroid and by scaling the considered exposure down to reach a number of counts of 180. For the most massive clusters, the exposure has to be scaled down to about 1~ks. The background level is thus much lower than the one measured in the low-S/N sub-sample. Therefore, once the exposure has been scaled down for the high-S/N sub-sample, we also add a scaled background realization (see \textsection \ref{sec:xray_sz}) to the event files in order to obtain S/N profiles that are similar to the ones measured for the low-S/N clusters. The 50 S/N profiles obtained with these new event files are shown along with the 17 original profiles of the low-S/N clusters in the right panel of Fig.~\ref{fig:snr_prof} (blue lines). The bimodality of the distribution of profiles shown in the left panel has been removed thanks to the re-scaling procedure. The mean S/N profile (black line) is higher than 2 in a radius range $20 \lesssim r \lesssim 400$~kpc. We will use the down-sampled event files of the high-S/N clusters only in \textsection \ref{sec:results} in combination with the original event files of the low-S/N clusters to estimate the redshift evolution of the fraction of cool-cores during cluster growth.


\section{Estimation of the ICM core properties}\label{sec:xray_sz}

\begin{figure*}[t]
\centering
\includegraphics[height=8cm]{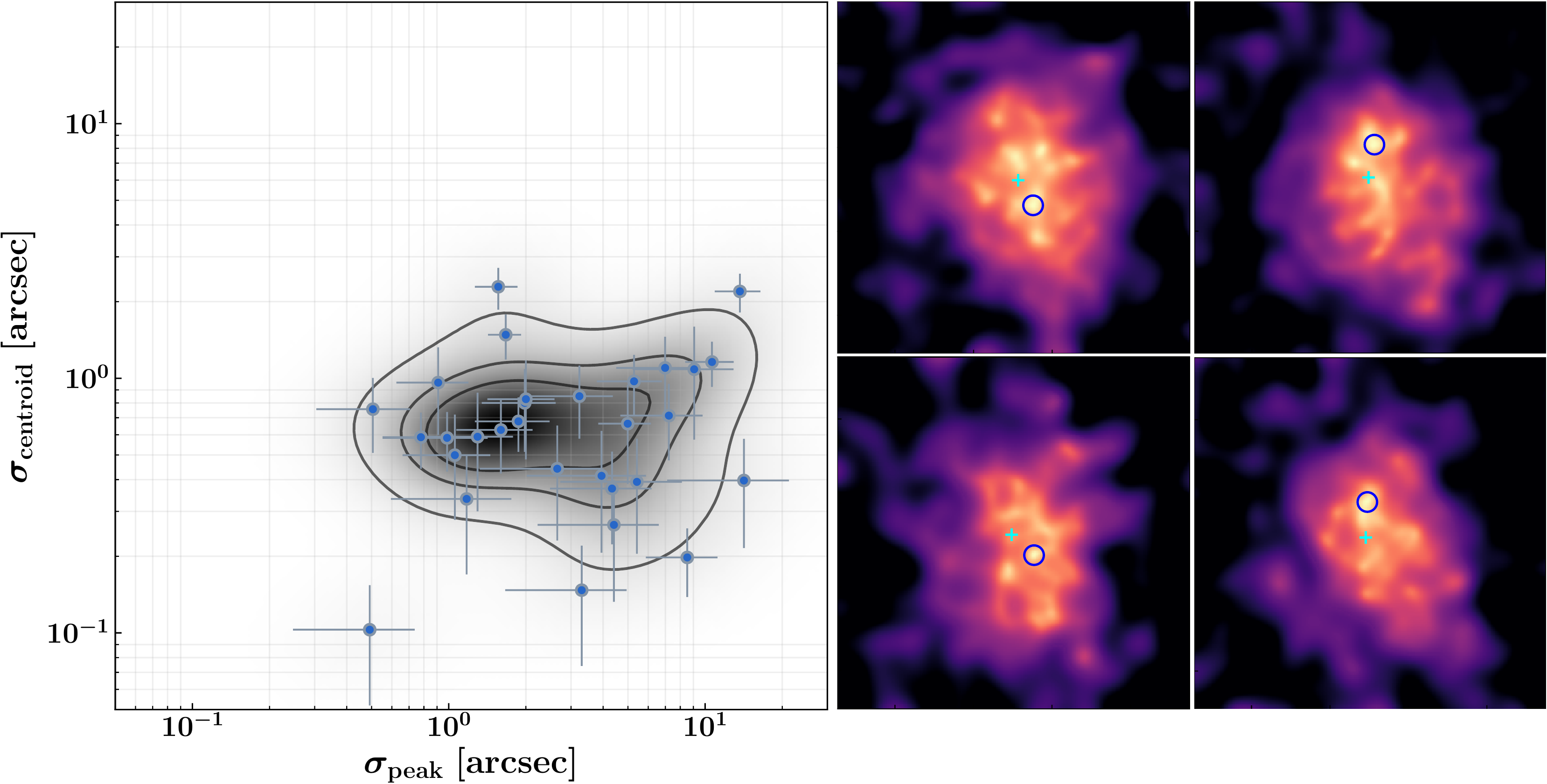}
\caption{{\footnotesize \textbf{Left:} Standard deviation of the angular distances measured between the centroid position obtained with all the counts and the ones estimated with down-sampled event files of 1000 counts as a function of the standard deviation obtained by considering the X-ray peak instead of the centroid for the 30 clusters in the high-S/N sub-sample with the highest S/N. The grey shaded region shows the Gaussian kernel-density estimate associated with the 2D distribution of points in the figure. Iso-density contours are shown in black at 10, 30, and 50\% of the peak amplitude of the distribution. \textbf{Right:} Four realizations of the \chandra\ image of SPT-CLJ0235-5121 obtained with 1000 counts out of the 4685 available in the 0.7-7~keV band. The images have been adaptively smoothed. The cyan cross shows the X-ray peak position obtained with all the counts while the ones measured in each image is shown with a blue circle.}}
\label{fig:peak_vs_cent}
\end{figure*}

This section presents the procedures used in order to estimate the ICM density profile of each cluster in the SPT progenitor sample. We first describe the pre-processing of the \chandra\ X-ray data which does not depend on the number of source counts. We then support our choice of the X-ray centroid as a deprojection center and further present the standard X-ray analysis used to estimate the ICM properties based on the original event files obtained for the high-S/N clusters. Finally, we present the joint X-ray/SZ analysis developed in order to push the investigation of the ICM density profile towards lower mass and higher redshift with the low-S/N sub-sample.

\subsection{X-ray data pre-processing}\label{subsec:preproc}

We have conducted the X-ray data reduction using the Chandra Interactive Analysis of Observations (CIAO) software v4.12 based on the calibration database (CALDB) v4.9.2 provided by the \chandra\ X-ray Center (CXC)\footnote{\url{https://cxc.cfa.harvard.edu/ciao/}}. Our main analysis steps follow the methodology described in \cite{mcd17} and references therein. The \texttt{chandra\_repro} script is used in order to reprocess the level 1 event files using the latest time-dependent gain adjustments and charge transfer inefficiency corrections. Flares are removed from lightcurves using the \texttt{lc\_clean} routine \citep{mar01}. We compute the exposure map associated with each observation in an energy band restricted from 0.7~keV to 7~keV with a center-band energy of 2.3 keV. We identify point sources using the \texttt{wavdetect} script based on a wavelet decomposition procedure \citep[\emph{e.g.}][]{vik98}. The resulting mask is used to produce a cleaned event file from which the X-ray surface brightness profile as well as the X-ray spectrum are extracted (see \textsection \ref{subsec:standard_X}). In the case of the low-S/N sub-sample, even though the clusters are detected at low-S/N, the observations are quite deep and are just as capable of detecting point sources as the observations realized for the high-S/N sub-sample. However, if we analyze a down-sampled event file from the high-S/N sub-sample (see \textsection \ref{subsec:rescale}), some point sources are not detected because of the low exposure considered to re-scale the S/N profile. Therefore, in this case, we consider the point source regions obtained by considering all the counts in the original event file in order to clean the down-sampled event file. Thus, we do not add any bias due to point source contamination in the analysis of the down-sampled event files.\\
\indent The X-ray background is defined as a combination of an instrumental background and an astrophysical background. The instrumental background is obtained through the normalization of unscaled stowed background files to the count rate observed in the 9-12~keV band. The astrophysical background is a combination of galactic foregrounds and the cosmic X-ray background. We estimate it using the ACIS-I chips regions that are free from cluster emission once both the particle background and point sources have been removed.\\The main result of the X-ray data pre-processing is the cluster surface brightness profile. It is defined as \citep{arn02}:
\begin{equation}
S_X(\theta) = \frac{1}{4\pi(1+z)^4} \, \epsilon(T,z)  \times EM(r)
\label{eq:XSB_def}
\end{equation}
where $z$ is the cluster redshift and $\epsilon(T,z)$ is the emissivity of the ICM at a temperature $T$ computed in the considered energy band. The cluster emission measure profile is given by:
\begin{equation}
\mathrm{EM}(r) = \int n_e n_p \, dl
\label{eq:em_prof}
\end{equation}
where $n_e$ and $n_p$ are the electron and proton density respectively, and $r = D_A(z)\theta$ is computed using the angular diameter distance at the cluster redshift $D_A(z)$.\\
Regardless of the available number of source counts, we follow the methodology introduced by \cite{mcd17} and extract the X-ray surface brightness profile using the \texttt{dmextract} routine in the 0.7-2.0~keV band in 20 annuli defined by:
\begin{equation}
r_{\mathrm{out},i} = (a+bi+ci^2+di^3)R_{500}~~i=1...20
\label{eq:XSB_an}
\end{equation}
where $(a,b,c,d) = (13.779, -8.8148, 7.2829, -0.15633) \times 10^{-3}$. This radial binning is optimized to allow us to efficiently sample the X-ray surface brightness profile of each cluster up to the highest redshifts considered in this study. The extracted surface brightness profiles are vignetting-corrected using the normalized exposure map estimated in the same energy band.

\subsection{Choice of deprojection center}\label{subsec:center}

\begin{figure}
\includegraphics[height=7cm]{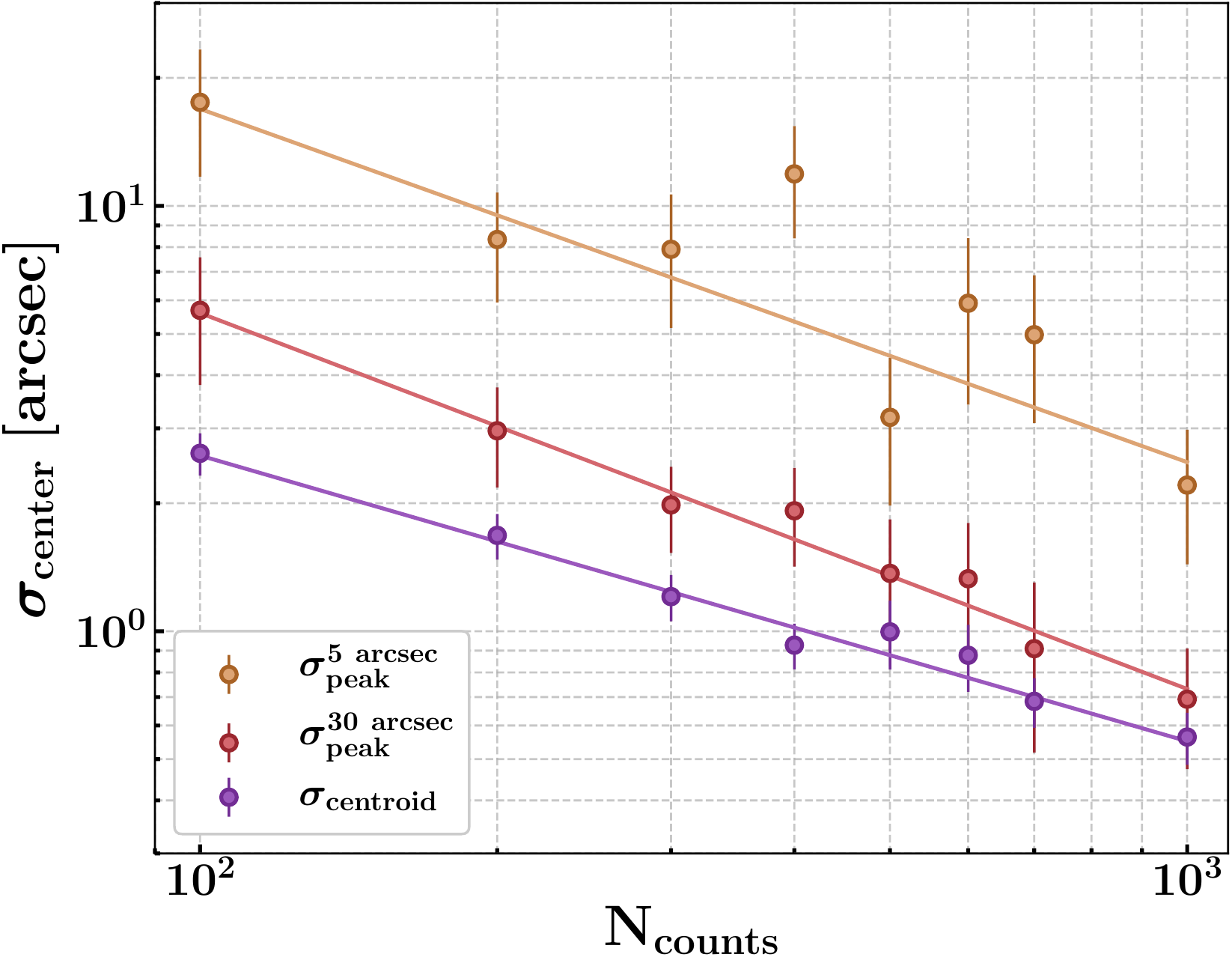}
\caption{{\footnotesize Standard deviation of the angular distances measured between the cluster center position obtained with all the counts and the ones estimated with down-sampled event files of $N_{\mathrm{counts}}$ counts as a function of $N_{\mathrm{counts}}$ for three different proxies of the cluster center based on the 30 clusters with the highest S/N in the sample. The results obtained for the X-ray peak estimated by smoothing the map with a 5~arcsec (30~arcsec) FWHM Gaussian kernel are shown in yellow (red). The results obtained for the X-ray centroid are shown in magenta. The computed values are fit with power laws shown with plane lines.}}
\label{fig:sig_evol_N}
\end{figure}

The choice of deprojection center considered for the extraction of the surface brightness profile can have a significant impact on the estimated core ICM properties in individual systems \citep[see \emph{e.g.}][]{mcd14,rup20}. We realize a dedicated analysis in order to find the most relevant definition of the deprojection center between the X-ray peak and the X-ray centroid to measure the surface brightness profile. The main driver of this choice is the stability of the location of the deprojection center with respect to S/N. We consider the complete event files of the 30 clusters from the progenitor sample with the highest S/N to realize this analysis. All these clusters are characterized with at least 1300 counts in the 0.7-7 keV band with an average number of counts of ${\sim}2700$. For each cluster in this sub-sample, we compute ten realizations of point source and background subtracted \chandra\ images with 1000 counts. We estimate the locations of the X-ray peak $P_i$ and the X-ray centroid $C_i$ for each realization. The position of the former is estimated by smoothing the X-ray map with a 5~arcsec Full Width at Half Maximum (FWHM) Gaussian kernel. The centroid is computed in a circular region with a R$_{500}$ radius and with a center estimated iteratively starting from the X-ray peak. We also compute the locations of the X-ray peak $P_{all}$ and centroid $C_{all}$ by using all the counts available for each cluster. We compute the standard deviation $\sigma_{\mathrm{peak}}$ of the angular distances between $P_{all}$ and $P_i$ from the ten realizations associated with each cluster in order to test the stability of the location of the X-ray peak across different observations of the same cluster with 1000 counts. We do the same analysis by considering the X-ray centroid in order to obtain $\sigma_{\mathrm{centroid}}$. The error bars associated with the measurements of $\sigma_{\mathrm{peak}}$ ($\sigma_{\mathrm{centroid}}$) are obtained by bootstrapping the estimates of the angular distances between $P_{all}$ ($C_{all}$) and $P_i$ ($C_i$) for each cluster.\\
\indent We show the estimates of $\sigma_{\mathrm{peak}}$ and $\sigma_{\mathrm{centroid}}$ for the considered sub-sample of clusters in the left panel of Fig.~\ref{fig:peak_vs_cent}. The Gaussian kernel-density estimate associated with the 2D distribution of measurements is shown in grey with black iso-density contours at 10, 30, and 50\% of the peak amplitude of the distribution. The distribution across the $\sigma_{\mathrm{peak}}$ axis is clearly skewed with an extended tail towards high values of $\sigma_{\mathrm{peak}}$. We measure standard deviations of the angular distances between $P_{all}$ and $P_i$ that are between 10 and 20~arcsec in few systems. These large values are observed in clusters with a disturbed core such as SPT-CLJ0235-5121 (see right panel of Fig.~\ref{fig:peak_vs_cent}). In such systems, the gas distribution is nearly flat in the core. The X-ray peak location is thus very sensitive to Poisson fluctuations of both the background and the ICM signal itself. However, the highest value of $\sigma_{\mathrm{centroid}}$ measured in this sub-sample is 2.3~arcsec. Half of the sub-sample is characterized by a $\sigma_{\mathrm{peak}}$ estimate that is larger than this value. This shows that the fraction of clusters in this sample with a poorly defined X-ray peak is non-negligible. On the other hand, the distribution of $\sigma_{\mathrm{centroid}}$ is quite symmetric, with a median value $\mathrm{med}(\sigma_{\mathrm{centroid}}) = 0.6~\mathrm{arcsec}$. The X-ray centroid location is thus very stable with respect to cluster dynamical state.\\
\indent We repeat the same analysis with different number of cluster counts in the down-sampled event files, from 100 to 1000 counts in the 0.7-7 keV band. For each analysis, we compute the mean value of $\sigma_{\mathrm{peak}}$ and $\sigma_{\mathrm{centroid}}$ based on the 30 values obtained for this sub-sample of clusters. We smooth the X-ray map with a 5~arcsec and a 30~arcsec FWHM Gaussian kernel before finding the location of maximum emission in both cases. We report the evolution of $\sigma_{\mathrm{peak}}^{5~\mathrm{arcsec}}$, $\sigma_{\mathrm{peak}}^{30~\mathrm{arcsec}}$, and $\sigma_{\mathrm{centroid}}$ as a function of the number of counts $N_{\mathrm{counts}}$ in Fig.~\ref{fig:sig_evol_N}. As expected, we find that these standard deviations decrease with increasing number of cluster counts in the X-ray map. The evolution of the standard deviations with cluster counts is well modeled by a power law (see plane lines in Fig.~\ref{fig:sig_evol_N}). Interestingly, we find that smoothing the X-ray map by a 30~arcsec FWHM Gaussian kernel (red points) is not sufficient to reach the stability of the centroid position (magenta points). This analysis shows that in the case of the low-S/N sub-sample, we expect a $\sigma_{\mathrm{centroid}}$ value of the order of 2~arcsec while the $\sigma_{\mathrm{peak}}^{5~\mathrm{arcsec}}$ one is five times larger. The X-ray centroid location is thus quite stable with respect to S/N.\\
\indent We note that there is one case in which considering the X-ray peak instead of the X-ray centroid is more appropriate to characterize the ICM properties. If a cool-core cluster has a core that is significantly offset with respect to its centroid \citep[see \emph{e.g.}][]{mcd14,rup20}, then choosing the centroid may induce a miss-classification of such a cluster as a system with a disturbed core. However, there are only two clusters out of 30 in this sub-sample that satisfy this condition. Therefore, following \cite{mcd13}, we will consider in this paper that a cool-core cluster is a system with an over-dense cool gas region located at its barycenter. Therefore, we choose to consider the X-ray centroid as a deprojection center in the following for its stability with respect to both the cluster core dynamical state and the observation S/N.

\subsection{Standard processing using X-ray spectroscopy}\label{subsec:standard_X}

The progenitor sample described in \textsection \ref{sec:spt_prog_samp} contains 50 clusters observed with enough S/N to enable estimating a mean ICM temperature through X-ray spectroscopy. This section presents the standard X-ray analysis that we have realized for these clusters in order to estimate their ICM density profile based on all the counts available in their respective event files.

\subsubsection{X-ray temperature}\label{subsubsec:temperature}

\begin{figure*}[t]
\centering
\includegraphics[height=6.5cm]{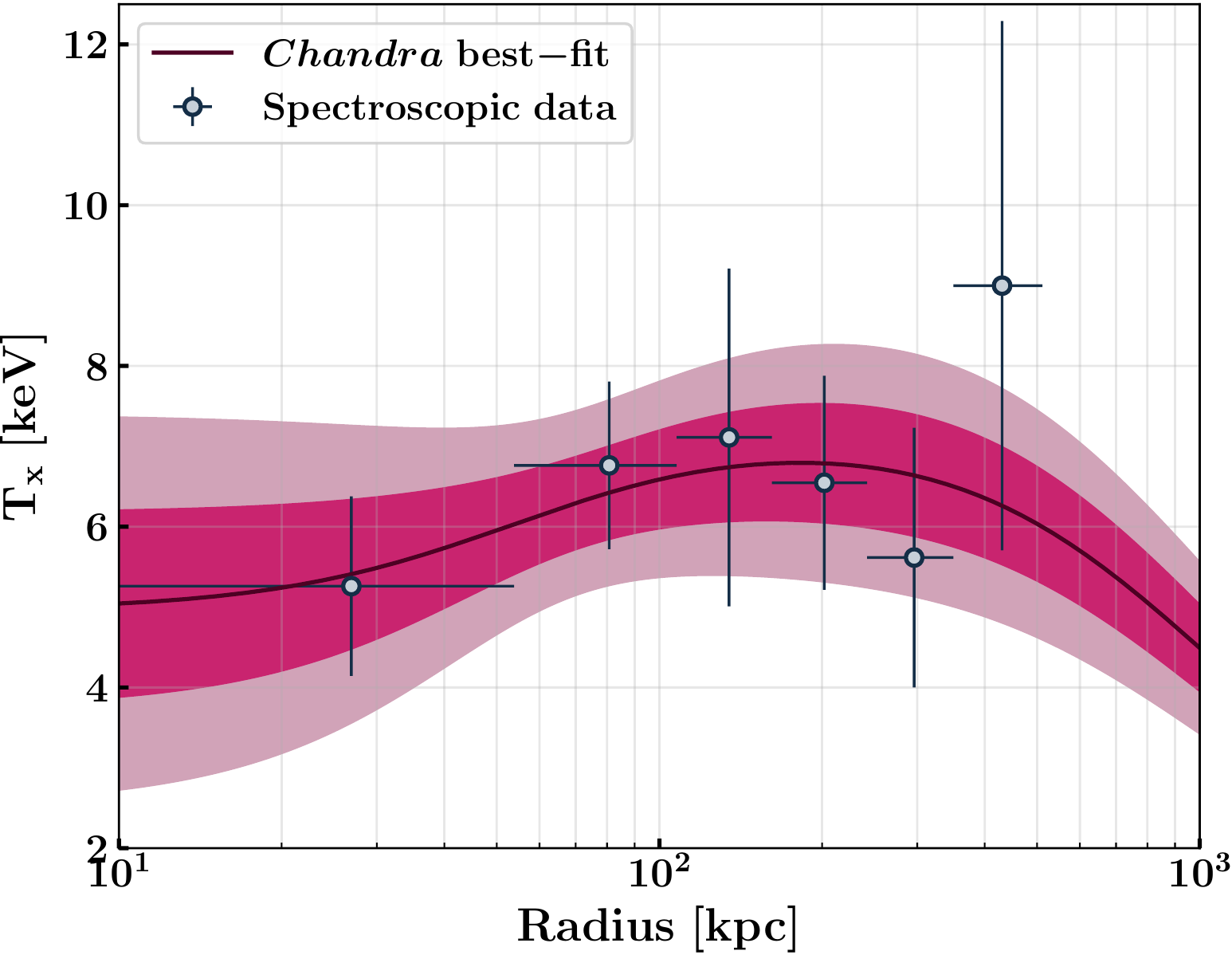}
\hspace{0.5cm}
\includegraphics[height=6.5cm]{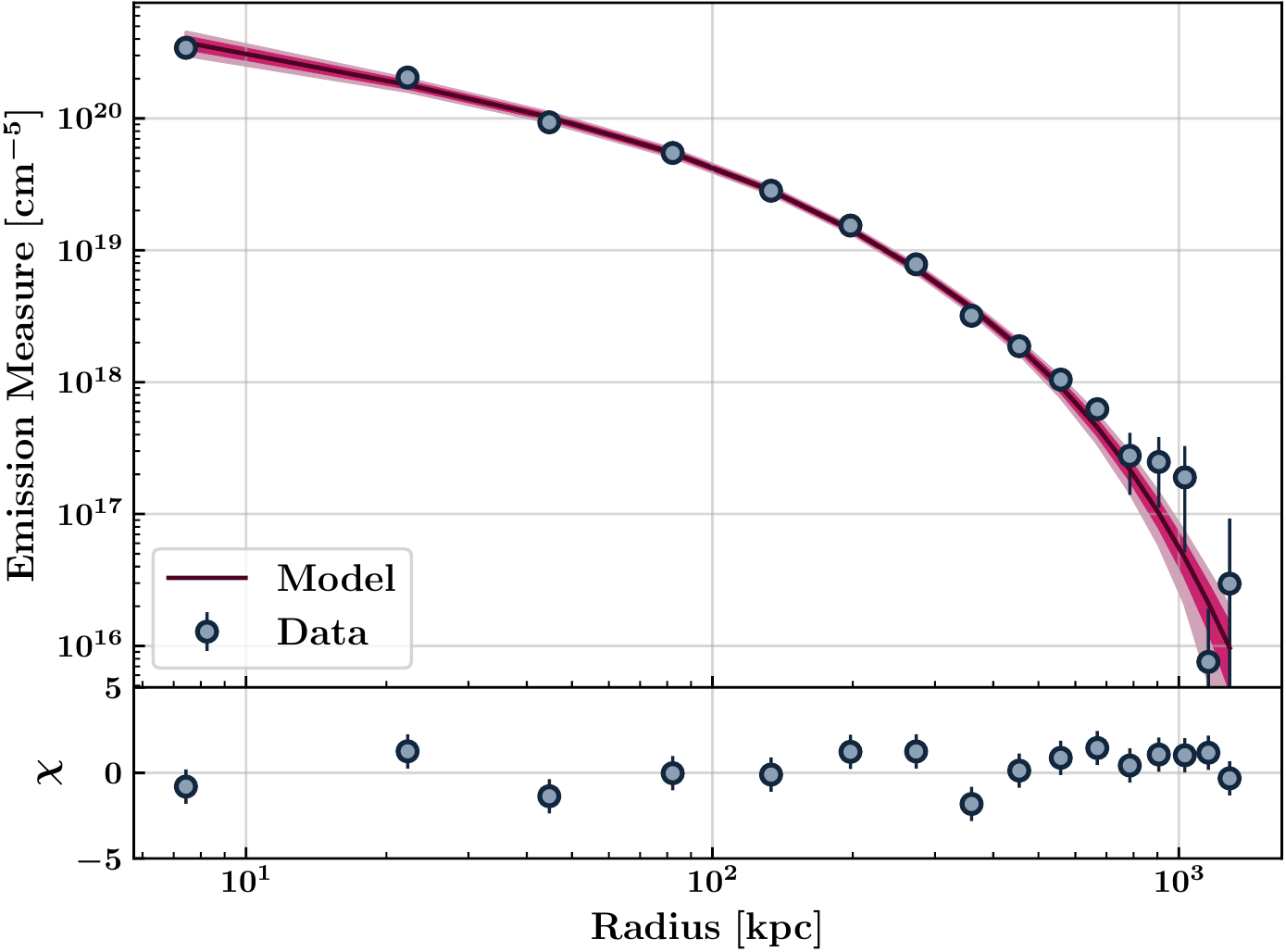}
\caption{{\footnotesize Temperature (left) and emission measure profile (right) of SPT-CLJ0304-4921 estimated from the standard X-ray analysis described in \textsection \ref{subsec:standard_X}. The spectroscopic temperatures estimated from the X-ray spectra and the emission measure computed from the X-ray surface brightness profile are shown with grey points. The red curves show the best-fit Vikhlinin temperature model and emission measure model respectively. In both panels, the red shaded areas show the $1\sigma$ and $2\sigma$ confidence regions. The significance of the residual between the data and the emission measure model is shown in the lower panel on the right.}}
\label{fig:standard_x}
\end{figure*}

We use the \texttt{specextract} script in circular annuli centered on the X-ray centroid (see \textsection \ref{subsec:center}) in order to extract X-ray spectra at different angular distances from the deprojection center for each cluster in the high-S/N sub-sample. We use a minimum of 500 counts in the 0.7-7.0~keV energy band for each spectrum. If the total number of counts available is lower than 1000, we use a core-excised annulus between $0.15$~R$_{500}$ and R$_{500}$ and extract a single spectrum to characterize the ICM. We also extract a background spectrum in the regions of the ACIS-I chips that are free from cluster emission. We subtract the spectrum of the particle background from all spectra by using stowed background files scaled to the number of counts observed in the 9-12~keV band. The spectra of the ICM signal and the astrophysical background are binned in order to obtain a S/N higher than 3 in each energy channel. The ICM spectra and the astrophysical background spectrum are jointly fit using CIAO's \emph{Sherpa} package. We use the \texttt{XSPEC} \citep{arn96} single-temperature plasma \citep[\texttt{APEC};][]{smi01} model to fit the cluster emission in combination with a soft X-ray Galactic background (\texttt{APEC}, $k_BT_X = 0.18~\mathrm{keV}$, $Z = Z_{\odot}$, $z=0$), a hard X-ray cosmic spectrum \texttt{BREMSS} with a fixed temperature $k_BT_X = 40~\mathrm{keV}$, and a Galactic absorption model (\texttt{PHABS}). We use the Galactic column density values found by \cite{kal05} at the latitude of each cluster. The cluster redshift is fixed in the cluster emission model to the value found in the updated version of the SPT-SZ catalog \citep{boc19} or in the SPTpol 100d catalog \citep{hua19}. As the iron emission line is usually poorly detected in each spectrum, we choose to fix the ICM metallicity to $Z = 0.3 Z_{\odot}$ \citep{man20}. The ICM spectroscopic temperature and the different model normalizations are allowed to vary during the fitting procedure. Therefore, at the end of this analysis, we obtain either a projected spectroscopic temperature profile or a mean spectroscopic temperature of the ICM.\\
\indent We fit the spectroscopic temperature estimates using the following temperature model:
\begin{equation}
T(x) = 1.35 \, T_{\mathrm{ICM}} \frac{(x/0.045)^{1.9} + \alpha}{(x/0.045)^{1.9} + 1} \frac{1}{(1+(x/0.6)^2)^{0.45}}
\label{eq:T_fit}
\end{equation}
where $x = r/\mathrm{R_{500}}$, $r$ is the projected radius, and $T_{\mathrm{ICM}}$ and $\alpha$ are free parameters giving the mean ICM temperature and the ratio between the core temperature and $T_{\mathrm{ICM}}$, respectively. If the number of temperature data points is too low to fit these two parameters (typically at least three points are needed), we fix $\alpha$ to the universal value found by \cite{vik06}, \emph{i.e.} $\alpha = 0.45$, and only fit the normalization of the profile.\\
As an example, we show the temperature profile model fit to the spectroscopic temperatures estimated at different radii from the X-ray centroid of SPT-CLJ0304-4921 in the left panel of Fig.~\ref{fig:standard_x} (dark red line). In this and subsequent figures, a specific cluster is chosen for clarity, but as a representative of the considered sample to illustrate an important step in our analysis pipeline. This fit is used to estimate the emissivity of the ICM in order to compute the emission measure profile from the X-ray surface brightness profile (see Eq.~\ref{eq:XSB_def} and \textsection \ref{subsubsec:density}).

\subsubsection{Density profile}\label{subsubsec:density}

We estimate the ICM electron density profile $n_e(r)$ from the cluster emission measure related to the measured X-ray surface brightness profile given in Eq.~(\ref{eq:XSB_def}). We first estimate the ICM emissivity by taking into account the effects of the Galactic absorption and the variations of {\chandra}'s effective area as a function of energy and position in the field of view. The emissivity is thus computed by estimating the normalization factor of the APEC model associated with the count rate $R$ measured in each annulus of area $A$ of the surface brightness profile\footnote{\url{https://heasarc.gsfc.nasa.gov/xanadu/xspec/manual/XSmodelApec.html}}:
\begin{equation}
\mathrm{norm}(R,A) = \frac{10^{-14}}{4\pi [D_A(z) (1+z)]^2} \int n_e n_p \, d\Omega dl
\label{eq:normalization}
\end{equation} 
where $dl$ and $d\Omega$ are the line of sight and solid angle differential elements respectively. The temperature of the APEC model is fixed in each annulus to the one given by the best-fit temperature profile associated with the spectroscopic measurements (see Eq.~\ref{eq:T_fit} and \textsection \ref{subsubsec:temperature}). Knowing both the temperature and redshift, and fixing the metallicity to a constant value $Z = 0.3 Z_{\odot}$ in each annulus, we thus compute the conversion coefficient between surface brightness and emission measure as a function of angular distance from the X-ray centroid. This allows us to measure the emission measure profile (see Eq.~\ref{eq:em_prof}) from the X-ray surface brightness profile.\\

We estimate the ICM electron density profile $n_e(r)$ from a Bayesian forward fit of the emission measure profile. We model the electron density distribution using a Vikhlinin parametric model \citep[VPM;][]{vik06}:
\begin{equation}
        n_e(r) = n_{e0}  \frac{ \left[\frac{r}{r_c} \right]^{-\frac{\alpha}{2}}}{\left[1+\left(\frac{r}{r_c}\right)^2 \right]^{\frac{3\beta}{2}-\frac{\alpha}{4}}}\frac{1}{\left[ 1+\left(\frac{r}{r_s}\right)^{\gamma} \right]^{\epsilon/2 \gamma}},
\label{eq:VPM}
\end{equation}
where $n_{e0}$ is the central density of the ICM and $r_c$ and $r_s$ are respectively the core radius and the transition radius at which an additional steepening characterized by a width $\gamma$ occurs in the profile. The parameters $\beta$ and $\epsilon$ give the inner and outer slopes of the profile, respectively. The slope of the power law-type cusp in the cluster core is given by $\alpha$. We do not have enough S/N in the core of most clusters in the progenitor sample to alleviate the degeneracy between the $\alpha$, $\beta$, and $n_{e0}$ parameters. Therefore, we choose to use the simplification introduced by \cite{mro09} and fix $\alpha$ to zero in the following. We also build upon the results of \cite{vik06b} and fix the $\gamma$ value to three. The ICM proton density profile $n_p(r)$ is computed from $n_e(r)$ by assuming an ionization fraction $n_e/n_p = 1.199$ associated with a fully ionized plasma with an abundance of $0.3 Z_{\odot}$ \citep{and89}. We note that, by using a parametric model to estimate an ICM quantity, the shape of the profile is constrained a priori by the number of degrees of freedom in the model compared to a non-parametric procedure. This leads by construction to smaller uncertainties on the final profile \citep{man11}.\\

The fitting procedure is based on a Markov chain Monte Carlo (MCMC) analysis based on the \texttt{emcee} python package \citep{for13} in order to efficiently sample the parameter space defined by the five free parameters $n_{e0}$, $r_c$, $\beta$, $r_s$, and $\epsilon$. We use the following Gaussian likelihood function in order to estimate the best-fit parameters of the VPM model for each cluster in the high-S/N sub-sample:
\begin{equation}
-2 \mathrm{ln} \, \mathscr{L}_{\mathrm{CXO}} = \sum_{i=1}^{N_{\mathrm{bin}}} [(EM_{CXO} - \widetilde{EM}) /  \Delta EM_{CXO}]^2_i
\end{equation}
where $N_{\mathrm{bin}}$ is the number of data points in the emission measure profile $EM_{CXO}$ estimated from the \chandra\ surface brightness profile using the conversion coefficient defined by Eq.~(\ref{eq:normalization}) and $\Delta EM_{CXO}$ are the associated error bars. The model of the emission measure profile is given by $\widetilde{EM}$. It is computed by integrating the product of the $n_e(r)$ and $n_p(r)$ profiles given by the VPM model along the line of sight (see Eq.~\ref{eq:em_prof}). We use 100 Markov chains and 10,000 steps to realize the MCMC analysis. We ensure the convergence of the chains by computing both the \cite{gel92} convergence test and the chain autocorrelation function in order to select independent samples in the posterior distribution. We sample the posterior distribution after a burn-in cutoff discarding the first quarter of samples in order to compute the best-fit electron density profile along with its associated uncertainties. We show the best-fit emission measure model of SPT-CLJ0304-4921 in the right panel of Fig.~\ref{fig:standard_x} (dark red line). The $1\sigma$ and $2\sigma$ confidence levels associated with the model are shown with dark and light red regions, respectively. As shown in the lower panel, we do not measure any deviation between the data and the model that is higher than $3\sigma$. We obtain similar results for all 50 clusters in the high-S/N sub-sample. 

\subsection{Joint X-ray/SZ analysis}\label{subsec:xsz_analysis}

The complementarity between \chandra\ observations and millimeter data offers a unique opportunity to probe cluster physics in the $1 < z < 2$ range. In this section, we demonstrate how combining X-ray and SZ observations can allow us to estimate the ICM density profile of the low-S/N clusters with relative uncertainties of the order of 20\%. We first motivate our choice to use SZ data in combination with X-ray observations to add constraining power on the ICM properties. Then, we detail the joint analysis procedure that we have developed to estimate the ICM density profile based on \chandra\ and SPT data.

\subsubsection{Motivation}\label{subsubsec:motivate}

\begin{figure*}[t]
\centering
\includegraphics[height=6.6cm]{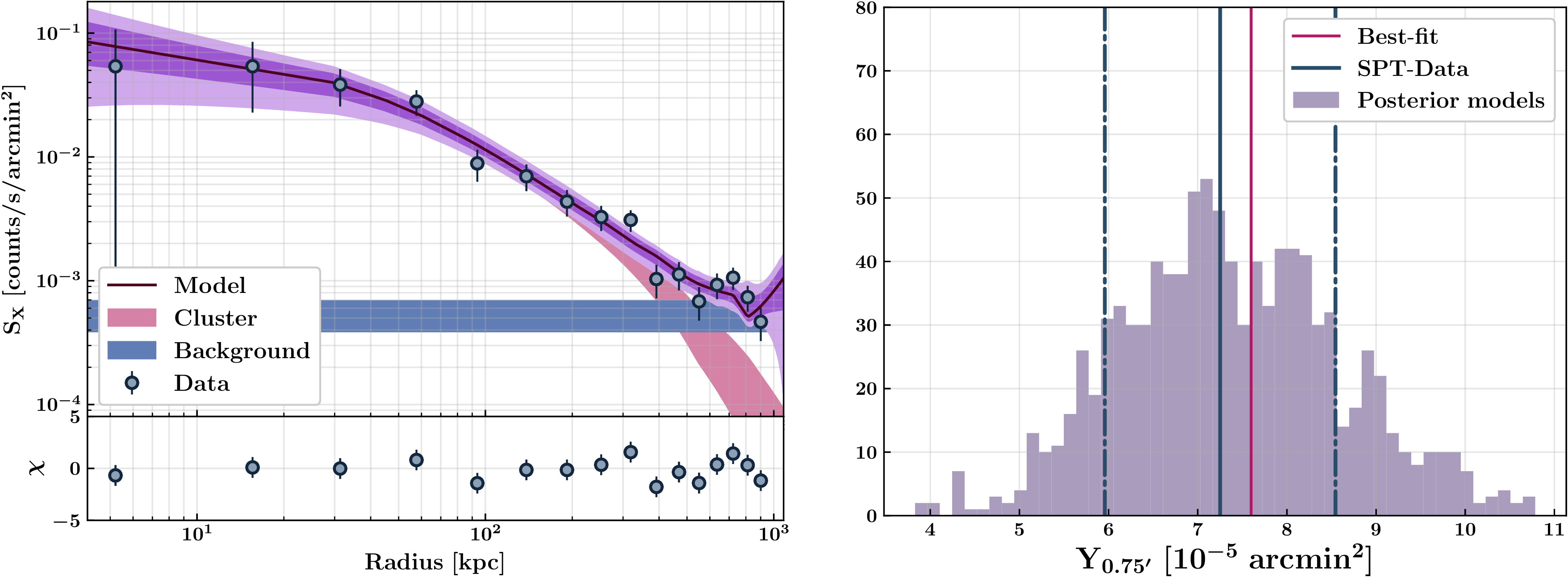}
\caption{{\footnotesize \textbf{Left:} Best-fit surface brightness profile (dark purple line) estimated from the joint X-ray/SZ analysis (see \textsection \ref{subsubsec:xsz_method}) of the \chandra\ data (grey points) measured on SPT-CLJ2343-5024 at $z = 0.88$. The $1\sigma$ and $2\sigma$ confidence levels on the best-fit profile are shown with the dark and light magenta regions, respectively. The red and blue regions give the $1\sigma$ intervals around the best-fit models of the ICM-induced and background X-ray emissions, respectively. The lower panel shows the difference between the data and the model, weighted by the measurement errors. \textbf{Right:} Posterior distribution of the integrated Compton parameter models of SPT-CLJ2343-5024 (grey) along with the best-fit value (red line). The measured SPT value and its associated uncertainties are shown with the solid and dash-dotted blue lines, respectively.}}
\label{fig:sx_ysz_fit}
\end{figure*}

The average number of counts available in the 0.7-2~keV band for the clusters in the low-S/N sub-sample is 180 (see Tab.~\ref{tab:low_sn} in Appendix~\ref{sec:app_A}). With such statistics, the X-ray spectrum extracted in a single annulus mapping the radius range $0.15 \mathrm{R}_{500} < r <  \mathrm{R}_{500}$ is fully compatible with a background-only spectrum (see Fig.~\ref{fig:spec_180} in Appendix~\ref{sec:app_B}). This leads to huge systematic uncertainties on the estimated emission measure profile with a standard X-ray analysis (see \textsection \ref{subsec:gain}). It is thus essential to consider additional information to constrain the ICM density profile of these clusters.\\
\indent The thermal Sunyaev-Zel'dovich effect \citep[SZ;][]{sun72,sun80} has already been shown to be an excellent probe to complement X-ray observations and push the investigation of the ICM properties to higher redshifts \citep[see \emph{e.g.}][]{ada15,rup17,rup18,rom18,ker20}. This effect is due to the inverse Compton scattering of CMB photons by energetic ICM electrons. It induces a variation of the apparent brightness of the cosmic microwave background (CMB) in the direction $\hat{n}$ of the sky for which amplitude is given by the Compton parameter:
\begin{equation}
        y_{SZ}(\hat{n}) = \frac{\sigma_{\mathrm{T}}}{m_{e} c^2} \int P_{e} \, dl,
        \label{eq:y_compton}
\end{equation}
where $\sigma_{\mathrm{T}}$ is the Thomson scattering cross section, $m_{e}$ the mass of the electron, $c$ the speed of light, and $P_{e}$ the electron pressure distribution of the ICM. The thermal SZ effect is thus a direct probe of the ICM pressure profile. The SZ surveys that have been realized so far lack the angular resolution to constrain the Compton parameter profile in the core of high redshift clusters. The SZ observable is thus given by the integrated Compton parameter:
\begin{equation}
Y_{SZ}^{\theta_{\mathrm{max}}} = 2\pi \int_0^{\theta_{\mathrm{max}}} y(\theta) \theta \,d\theta
\label{eq:int_Y}
\end{equation}
where $y(\theta)$ is the Compton parameter profile given by Eq.~(\ref{eq:y_compton}) and $\theta_{\mathrm{max}}$ is the maximum angular distance from the cluster center considered to integrate the SZ signal.\\
\indent As the ICM is well described by an ideal gas, the ICM pressure is given by $P_e = n_e \times k_B T_e$. Therefore, it is possible to break the degeneracy between the ICM density $n_e$ and temperature $T_e$ in the X-ray surface brightness profile (see Eq.~\ref{eq:XSB_def} and \ref{eq:em_prof}) by solving the following system of two equations in two unknowns $n_e$ and $T_e$:
\begin{equation}
\begin{cases}
S_X = \frac{1}{4\pi(1+z)^4} \, \epsilon(T_e,z) \int \frac{n_e^2}{1.199} \, dl \\
Y_{SZ}^{\theta_{\mathrm{max}}} = 2\pi \frac{k_B\sigma_{\mathrm{T}}}{m_{e} c^2} \int_0^{\theta_{\mathrm{max}}} \int n_eT_e \, \theta \, dl \, d\theta
\end{cases}
\label{eq:syst2eq}
\end{equation}
where the left hand side of each equation is a quantity that is directly measured from the X-ray and SZ observations.\\
\indent We emphasize that solving the system given by Eq.~(\ref{eq:syst2eq}) does not require access to high angular resolution SZ observations. The integrated Compton parameter is already provided by most SZ cluster catalogs \citep{pla16b,ble15,has13} and can be used directly in combination with X-ray observations in order to add constraining power on the models of the ICM thermodynamic properties.

\subsubsection{Analysis procedure}\label{subsubsec:xsz_method}

We consider the \chandra\ X-ray surface brightness profile extracted from the point source subtracted event files in the 0.7-2~keV band (see \textsection \ref{subsec:preproc}) and the SPT integrated Compton parameter $Y_{SZ}^{0.75'}$ integrated up to an angular distance of $0.75$~arcmin to solve Eq.~(\ref{eq:syst2eq}) for each cluster in the low-S/N sub-sample. The SPT integrated Compton parameter $Y_{SZ}^{0.75'}$ is only publicly available in the SPT-SZ catalog. Therefore, the same tool described in \cite{ble15} has been used in order to measure it for each cluster in the SPTpol 100d catalog \citep{hua19}. We do not subtract the background from the \chandra\ surface brightness profile as the Gaussian approximation cannot be considered with such shallow X-ray observations. Therefore, we decide to model the total X-ray surface brightness profile as $\hat{S_X} = S_X + B_X$ where $B_X$ is a constant background value.\\
\indent We model the ICM electron density profile using the VPM profile defined in Eq.~(\ref{eq:VPM}). The temperature model in Eq.~(\ref{eq:syst2eq}) is given by $k_B T_e = P_e / n_e$ where the electron pressure distribution is modeled by a generalized Navarro-Frenk-White model \citep[gNFW;][]{nag07}, given by:
\begin{equation}
        P_e(r) = \frac{P_0}{\left(\frac{r}{r_p}\right)^c \left(1+\left(\frac{r}{r_p}\right)^a\right)^{\frac{b-c}{a}}},
\label{eq:gNFW}
\end{equation}
where $c$ and $b$ are the inner and the outer slopes of the profile, $a$ defines the width of the transition occurring at the characteristic radius $r_p$, and $P_0$ is a normalization constant.\\
\indent As the ICM temperature is not known a priori we need to tabulate the values of the emissivity $\epsilon(T_e,z)$ in each annulus of the X-ray surface brightness profile as a function of temperature. Thus, prior to each analysis, we save the values of the conversion coefficients between X-ray surface brightness and emission measure (see Eq.~\ref{eq:normalization} and \textsection \ref{subsubsec:density}) for different ICM temperatures ranging from 0.1~keV to 30.1~keV with 1~keV steps in each annulus of the \chandra\ surface brightness profile.\\
\indent At each step of the analysis, we use the temperature model defined by Eq.~(\ref{eq:VPM}) and (\ref{eq:gNFW}) to interpolate the tabulated conversion coefficients between X-ray surface brightness and emission measure. We multiply the estimated conversion profile by the emission measure profile obtained by integrating the square of the VPM model along the line of sight in order to model the X-ray surface brightness profile induced by the ICM. We further add a constant background value $B_X$ to the profile to obtain the total X-ray surface brightness model. Furthermore, we integrate the gNFW model along the line of sight in order to estimate the Compton parameter profile. The latter is integrated up to an angular distance of $0.75$~arcmin to obtain the integrated Compton parameter model.\\
We compare the model to the data using the following likelihood function:
\begin{equation}
\begin{split}
\mathrm{ln} \, \mathscr{L}_{\mathrm{X/SZ}} =\, & \mathrm{ln} \, \mathscr{L}_{\mathrm{X}} + \mathrm{ln} \, \mathscr{L}_{\mathrm{SZ}} \\
=\, & \sum_{i=1}^{N_{X}} \left[ D_i^{X} \mathrm{ln}(M_i^{X}) - M_i^{X} -\mathrm{ln}(\Gamma(D_i^{X}+1))\right] \\
 & -\frac{1}{2} \left[ \frac{D^{SZ} - M^{SZ}}{\Delta D^{SZ}} \right]^2
\end{split}
\label{eq:xsz_likelihood}
\end{equation}
where $\mathrm{ln} \, \mathscr{L}_{\mathrm{X}}$ is the Poisson likelihood function associated with the \chandra\ surface brightness profile $D^{X}$ and the associated model $M^{X}$, both containing $N_{X}$ bins. It is essential to use a Poisson likelihood function in the regime of low S/N as the Gaussian approximation made in \textsection \ref{subsec:standard_X} is no longer valid. The Gaussian likelihood function $\mathscr{L}_{\mathrm{SZ}}$ compares the SPT measurement of the integrated Compton parameter $D^{SZ}$ associated with an uncertainty $\Delta D^{SZ}$ to the model $M^{SZ}$ obtained by integrating the gNFW profile.\\
\begin{table}
\begin{center}
{\footnotesize
\begin{tabular}{ccc}
\hline
\hline
Parameters & Min & Max \\
\hline
$n_{e0}$ & $0$ & $+\infty$ \\
$r_c$ & $0$ & $1000~\mathrm{kpc}$ \\
$\beta$ & $-\infty$ & $+\infty$ \\
$r_s$ & $0$ & $5000~\mathrm{kpc}$ \\
$\epsilon$ & $0$ & $+\infty$ \\
$P_0$ & $0$ & $+\infty$ \\
$r_p$ & $0$ & $1000~\mathrm{kpc}$ \\
$a$ & $0$ & $5$ \\
$b$ & $2$ & $20$ \\
$c$ & $0$ & $1.1$ \\
$B_X$ & $0$ & $+\infty$ \\
\hline
\end{tabular}}
\end{center}
\caption{Interval boundaries defining the uniform priors associated with the 11 free parameters considered in the MCMC analysis detailed in \textsection \ref{subsubsec:xsz_method}.}
\label{tab:priors}
\end{table}
We use a MCMC analysis in order to sample the parameter space defined by the five free parameters of the VPM model (see \textsection \ref{subsubsec:density}), the five free parameters of the gNFW model (see Eq.~\ref{eq:gNFW}), and the free background value associated with the \chandra\ surface brightness profile. We initialize the gNFW parameters to the universal pressure profile values found by \cite{arn10} using the SPT M$_{500}$ mass to normalize the model. We initialize the VPM parameters so that the combination of the VPM and the gNFW models gives a temperature model that is compatible within 50\% to the Vikhlinin temperature model (see Eq.~\ref{eq:T_fit}) normalized to the mean ICM temperature derived from the core-excised $M_{500}{-}T_X$ scaling relation of \cite{bul19}. The X-ray background parameter is initialized to the value found in the ACIS-I chips that are free from cluster emission. We use an additional scatter of 50\% to initialize these parameters on 400 different chains and run the MCMC for 10,000 steps in order to find the best-fit values of all 11 parameters.\\
\begin{figure*}[t]
\centering
\includegraphics[height=5.9cm]{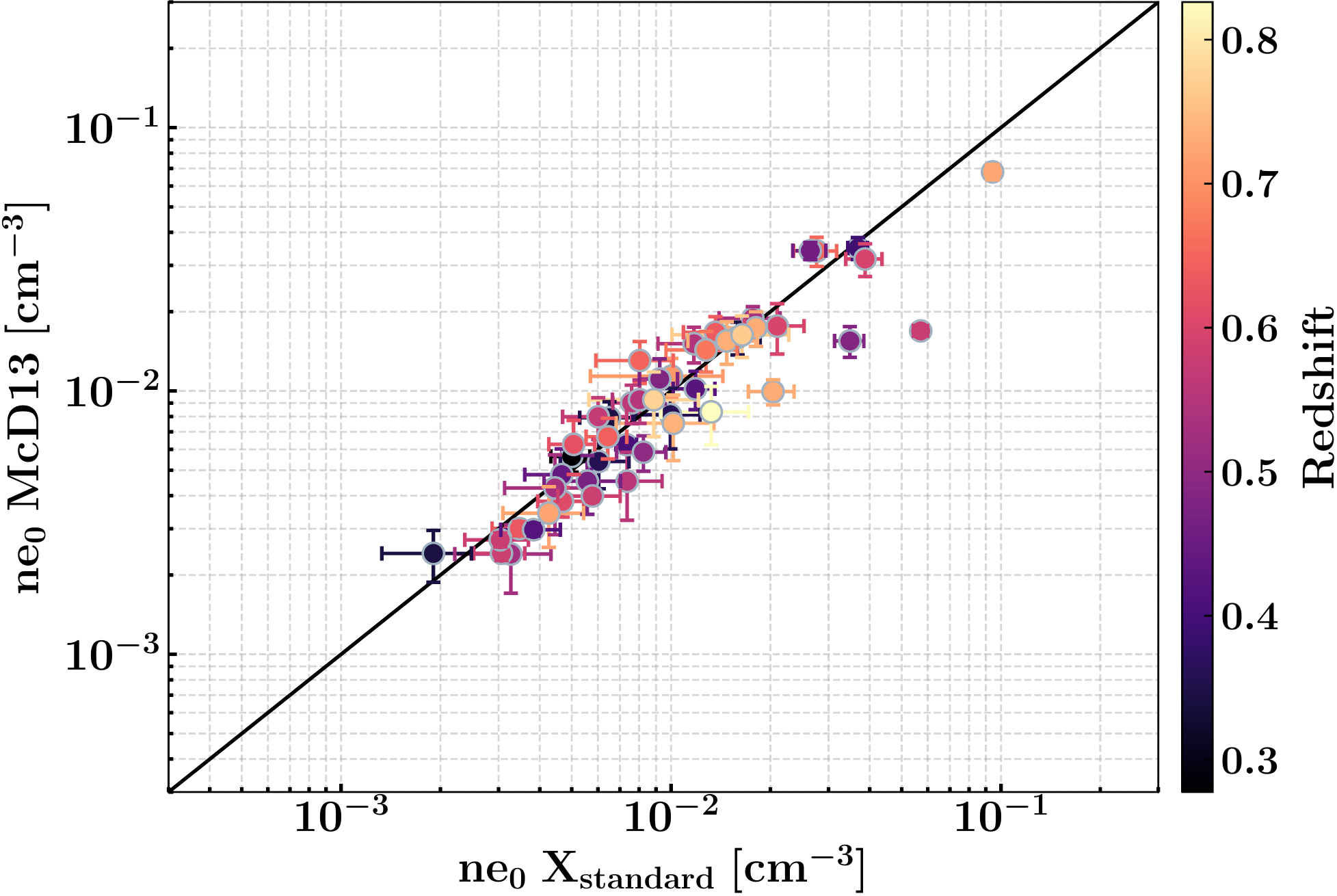}
\hspace{0.2cm}
\includegraphics[height=5.9cm]{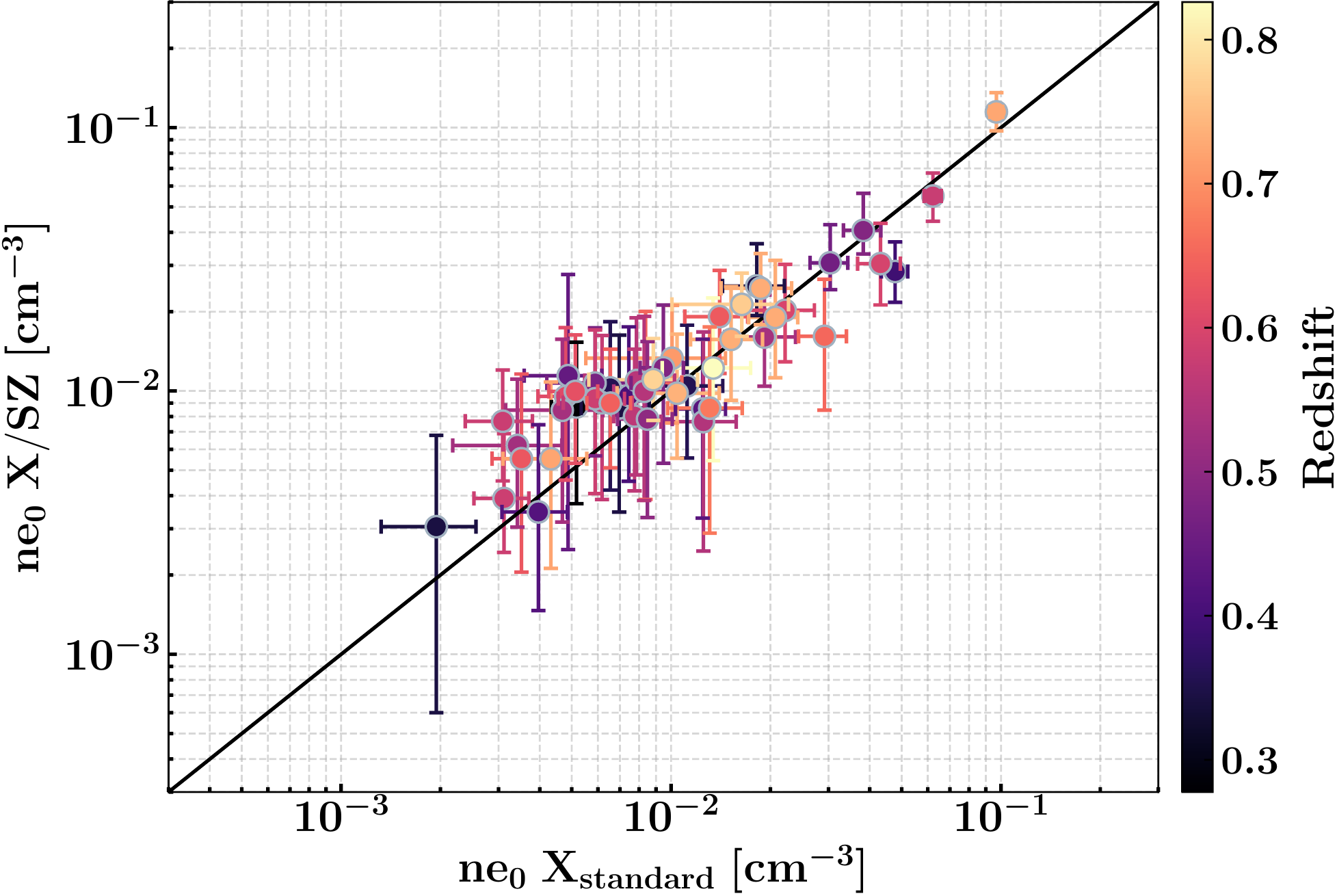}
\caption{{\footnotesize \textbf{Left:} Comparison between the central densities obtained by using all the X-ray counts available in the standard X-ray analysis described in \textsection \ref{subsubsec:density} and the ones estimated by \cite{mcd13} for the 49 XVP clusters in the high-S/N sub-sample. \textbf{Right:} Comparison between the central densities obtained with the standard X-ray analysis using all the counts and the joint X-ray/SZ analysis detailed in \textsection \ref{subsubsec:xsz_method} based on the down-sampled event files generated with the procedure described in \textsection \ref{subsec:rescale}. In both panels, the color of the points gives the redshift of the considered clusters in the high-S/N sub-sample and the black line is the line of equality.}}
\label{fig:validation}
\end{figure*}
We use uniform priors in combination with the likelihood function defined in Eq.~(\ref{eq:xsz_likelihood}) to estimate the posterior distribution. The prior boundaries are given in Tab.~\ref{tab:priors}. We consider physical boundaries as well as the results from \cite{pla13} to set these intervals. In addition to these uniform priors, we also consider physical constraints on the radial distributions of ICM thermodynamic properties to sample the parameter space. At each step of the MCMC analysis, we use the VPM and gNFW models in order to compute the hydrostatic mass profile $M_{\mathrm{HSE}}(r)$ and the ICM entropy profile $K_e(r)$ from the following equations:
\begin{equation}
\begin{split}
M_{\rm HSE}(r) \, & = -\frac{r^2}{\mu_{gas} m_p n_e(r) G} \frac{dP_e(r)}{dr} \\
K_e(r) \, & =  \frac{P_e(r)}{n_e(r)^{5/3}},
\end{split}
\end{equation}
where $m_p$ is the mass of the proton, $G$ is the gravitational constant, and  the mean molecular weight of the gas is given by $\mu_{\rm{gas}} = 0.61$. We then compute the radial derivatives of both the hydrostatic mass and the entropy profiles. A negative slope of the hydrostatic mass profile at a given radius $r+\Delta r$ would imply that negative mass has been added to the one contained in a sphere of radius $r$. Furthermore, in the presence of non-radiative processes, if clumps of low-entropy gas lie at equipotential lines that are above high-entropy gas, they will always sink to lower equipotential lines in a free-fall time thus making the entropy profile monotonically increasing \citep{toz01}. A shallower entropy slope has been observed in clusters where non-gravitational cooling and heating mechanisms are non-negligible compared to this gravitational process but the entropy profile remains nonetheless monotonically increasing \citep{voi05b}. Therefore, we require the slopes of both the hydrostatic mass and the entropy profiles to be non-negative between $10$~kpc and $1.5$R$_{500}$ in the MCMC sampling of the parameter space.\\
\indent We apply the same procedure described in \textsection \ref{subsubsec:density} in order to ensure the convergence of the chains and to compute the best-fit VPM profile and its associated uncertainties. We show the best-fit surface brightness profile of the low-S/N cluster SPT-CLJ2343-5024 at $z = 0.88$ characterized by 146 \chandra\ counts in the 0.7-2~keV band in the left panel of Fig.~\ref{fig:sx_ysz_fit} (dark purple line). The $1\sigma$ and $2\sigma$ confidence regions are shown with dark and light magenta regions, respectively. The significance of the residuals between the data (grey points) and the model is shown in the lower panel. We do not observe any residual with a significance larger than $3\sigma$ for all the clusters in the progenitor sample. We note that the background model (blue region) starts to be dominant over the ICM induced surface brightness profile (red region) at radii $r \gtrsim 500$~kpc. This is consistent with the lack of S/N observed at these radii in Fig.~\ref{fig:snr_prof}. We show a distribution of 1,000 values of the integrated Compton parameter $Y_{SZ}^{0.75'}$ estimated from the final posterior distribution in the right panel of Fig.~\ref{fig:sx_ysz_fit}. The central value and uncertainties extracted from the SPT data are shown with the solid and dash-dotted blue lines. The best-fit model of the integrated Compton parameter is shown with the red line. For all 67 clusters in the progenitor sample, the best-fit model for the integrated Compton parameter is always consistent with the SPT measurements.

\section{Performance of the joint analysis}\label{sec:performance}

The joint analysis described in \textsection \ref{subsubsec:xsz_method} needs to be validated in order to identify any systematic effect on the ICM density profiles estimated with this procedure. Furthermore, it is important to compare the results obtained with this method and the ones issued from the standard X-ray analysis detailed in \textsection \ref{subsubsec:density} in order to evaluate the information gain brought by the addition of SZ data in the procedure.

\subsection{Validation at $z < 0.8$}\label{subsec:validation}

\begin{figure*}[t]
\centering
\includegraphics[height=4.9cm]{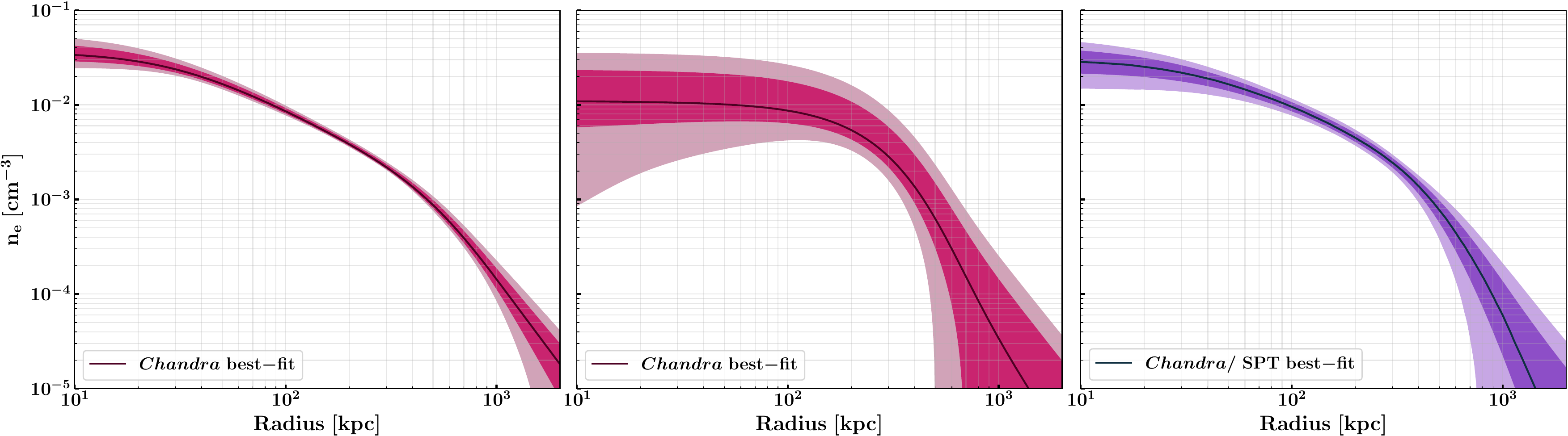}
\caption{{\footnotesize Density profiles of SPT-CLJ0304-4921 obtained with a standard X-ray analysis based on 2298 counts (left), with the same analysis based on 167 counts (middle), and with the joint X-ray/SZ analysis described in Sect. \ref{subsec:xsz_analysis} based on 167 counts and the SPT integrated Compton parameter (right). In each panel, the solid line shows the best-fit estimate of the ICM density profile and the dark and light colored regions give the $1\sigma$ and $2\sigma$ confidence levels, respectively.}}
\label{fig:ne_compare}
\end{figure*}

We validate the joint X-ray/SZ analysis by considering the high-S/N sub-sample. Indeed, the \chandra\ number of counts for each cluster in this sub-sample is sufficient to estimate the ICM density profile  with a standard X-ray analysis with good confidence levels (see \textsection \ref{subsubsec:density}).\\
\indent Thus, for each cluster in the high-S/N sub-sample, we run two different analyses. The first one considers all the counts available in the \chandra\ event file and is based on the standard procedure described in \textsection \ref{subsubsec:density}. The second one is based on the down-sampled event file (see \textsection \ref{subsec:rescale}) and it uses the joint analysis presented in \textsection \ref{subsubsec:xsz_method}. At the end of each analysis, we obtain a best-fit ICM density profile associated with each cluster at $z < 0.8$ and its associated uncertainties. We use these profiles in order to estimate the ICM core density at $10$~kpc for each cluster in both cases.\\
\indent We first compare the results estimated with the standard X-ray analysis with those obtained by \cite{mcd13} on the same clusters with a similar analysis in order to validate our standard X-ray pipeline. The comparison is shown in the left panel of Fig.~\ref{fig:validation}. We do not find any significant systematic deviation from the identity line (black) between our results and the ones estimated by \cite{mcd13}. Few outliers are identified. These systems all correspond to clusters with a disturbed morphology and a well-defined core. For example, the biggest outlier is SPT-CLJ2331-5051 which is a double peaked system characetrized by an angular separation of 2.8~arcmin between the two merging halos. As the X-ray centroid positions are estimated independently in our work and in the \cite{mcd13} analysis, the differences on the recovered ICM core densities to the high density end can be explained by a slight difference in the deprojection center locations considered in each analysis. Nevertheless, the average deviation between our results and the ones from \cite{mcd13} is consistent with zero in the whole dynamic range of core densities going from ${\sim}2\times 10^{-3}$ to $10^{-1}~\mathrm{cm}^{-3}$. Therefore, we consider that the standard X-ray processing described in \textsection \ref{subsubsec:density} is valid in the following.\\
\indent We compare the ICM core densities obtained at $10$~kpc with the standard X-ray analysis and with the joint X-ray/SZ analysis in the right panel of Fig.~\ref{fig:validation}. Although the available number of counts in the joint X-ray/SZ analysis is on average seven times lower than the one used in the standard X-ray analysis, we do not find any significant systematic deviation from the identity line (black) between the two estimates of the core density. The distribution of the ratio of the two estimates is however much more scattered than the one presented in the left panel of Fig.~\ref{fig:validation} especially at low core densities. The main driver of this increased scatter is photon statistics. In particular, some down-sampled event files do not have any count in the inner bin of the surface brightness profile computed in the 0.7-2~keV band (see Eq.~\ref{eq:XSB_an}). For these clusters, we use upper limits on the X-ray surface brightness in the inner bin in order to fit the ICM density profile following the procedure detailed in \textsection \ref{subsubsec:xsz_method}. This explains the origin of the small positive bias on the ICM core densities estimated in low core density clusters. We note however that this effect is taken into account in the uncertainties. Thus, this deviation with the line of equality is not significant.\\
\indent We further study any redshift dependent systematic effect by showing the redshifts associated with each cluster in the high-S/N sub-sample using a color scale in Fig.~\ref{fig:validation}. We do not find any redshift-dependent bias in either of the two panels. This implies that the redshift evolution of the angular size of the cluster core does not significantly impact our ability to recover the ICM core density with a joint X-ray/SZ analysis. As the joint X-ray/SZ analysis allows us to recover the ICM core densities over the same dynamic range covered by the estimates obtained with a standard X-ray analysis without significant bias, we consider that the joint X-ray/SZ pipeline is valid and can be used to estimate the ICM core densities in the low-S/N sub-sample.

\subsection{Gain in constraining power}\label{subsec:gain}

We run a standard X-ray analysis based on the down-sampled event files of the high-S/N sub-sample in order to evaluate the gain in constraining power brought by the joint X-ray/SZ analysis on the ICM density profile. As detailed in \textsection \ref{subsec:standard_X}, we first need to extract a spectrum from the down-sampled event files in order to estimate the mean ICM temperature. The latter is essential to convert the X-ray surface brightness profile into an emission measure profile that can be used to estimate the ICM density distribution. As explained in \textsection \ref{subsubsec:motivate}, these spectra are compatible with background only spectra in a large energy interval. The ICM temperatures estimated by fitting such spectra are associated with relative uncertainties of the order of 100\% and are usually compatible with zero. For this reason, there is a huge systematic uncertainty on the corresponding emission measure profiles. In particular, as the spectroscopic temperature tends towards zero, the emission measure tends towards infinity for a non-zero surface brightness (see Eq.~\ref{eq:XSB_def}). In practice, we set a minimum boundary for the ICM temperature of $0.5$~keV to ensure the plasma to be X-ray emitting. This ensures the systematic uncertainty associated with the emission measure profile to be finite.\\
\indent On the other hand, the median relative uncertainty on the SPT integrated Compton parameter is 25\%. For a given density model, this drastically limits the uncertainty on the associated temperature model needed to compute the emission measure profile. Thus, at each step of the joint X-ray/SZ MCMC analysis, the uncertainty on the emission measure profile is dominated by the Poisson fluctuations of the surface brightness profile and not by the systematic uncertainty induced by the lack of constraints on the temperature profile.\\ 
\begin{figure*}[t]
\centering
\includegraphics[height=6.5cm]{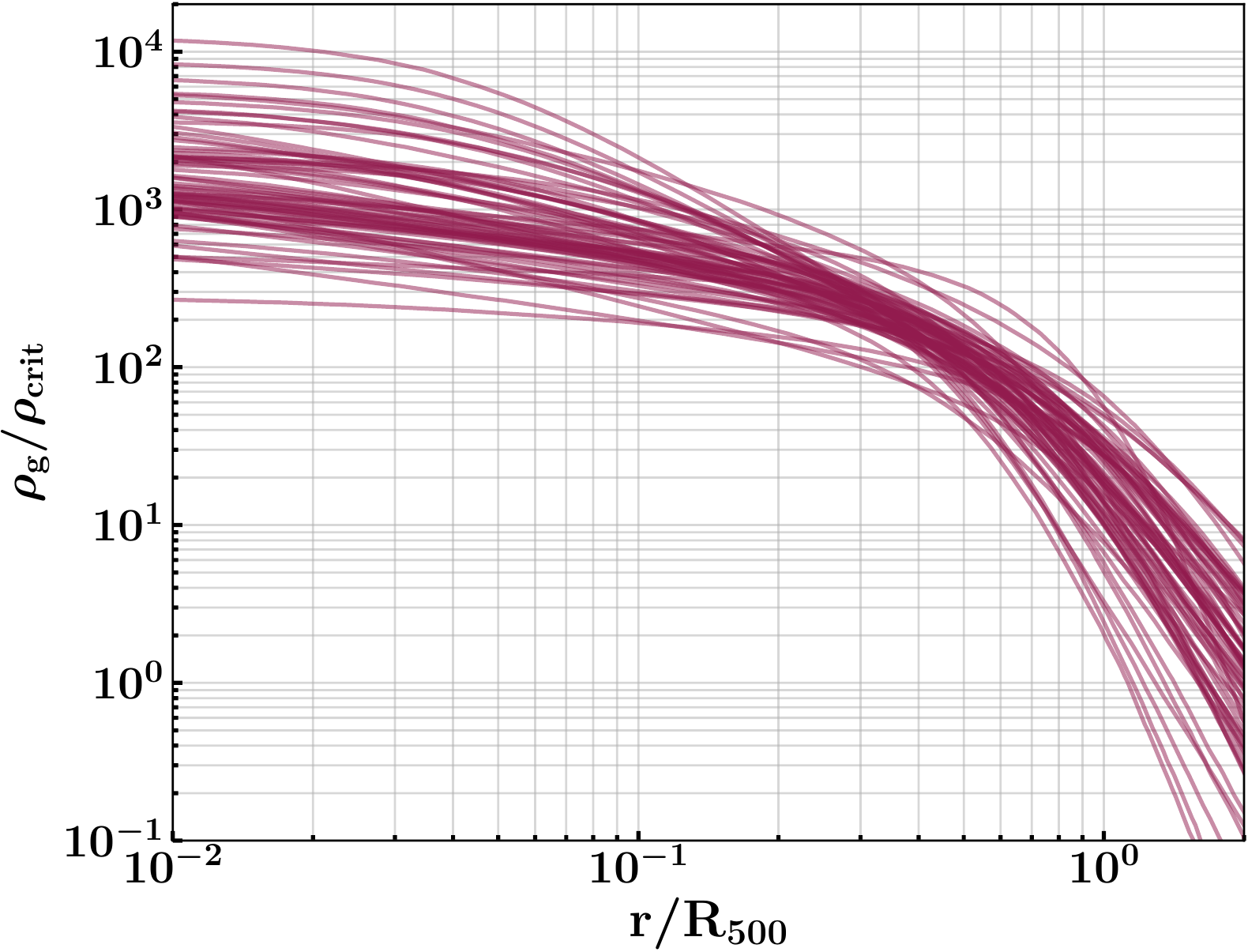}
\hspace{0.5cm}
\includegraphics[height=6.5cm]{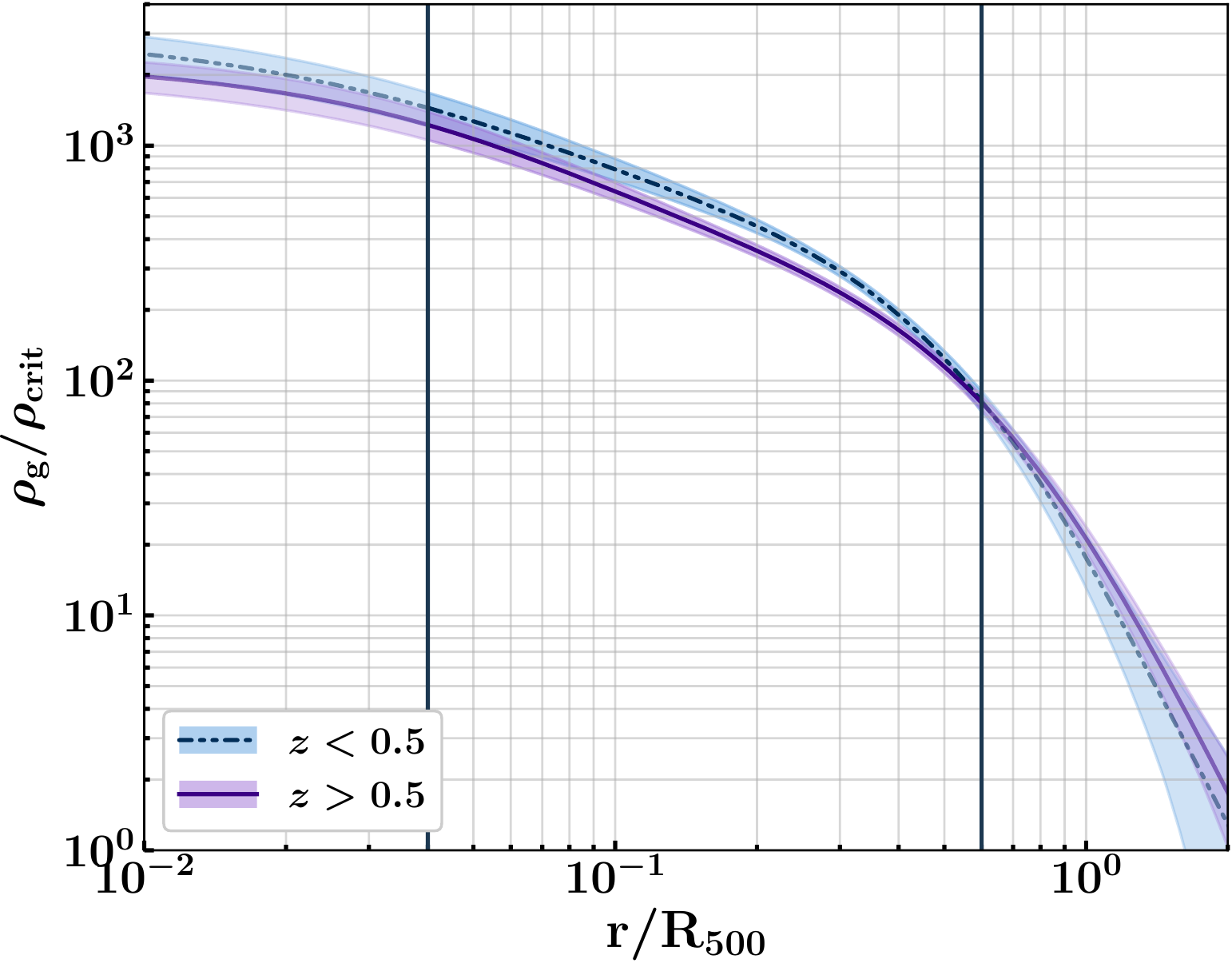}
\caption{{\footnotesize \textbf{Left:} Density profiles of the 67 clusters estimated from the joint X-ray/SZ analysis and scaled by the critical density of the universe at the considered redshifts. The radius is scaled by R$_{500}$ for each cluster. \textbf{Right:} Mean density profiles obtained at low (blue) and high (purple) redshift using the individual density profiles shown in the left panel. The shaded regions give the $1\sigma$ uncertainty on the mean profiles. The vertical solid lines delimit the inner and outer regions where the profiles are not constrained by the \chandra\ data and are thus extrapolated.}}
\label{fig:ne_profs}
\end{figure*}
We show how these effects translate into important gains on the relative uncertainties associated with the ICM density profile in Fig.~\ref{fig:ne_compare}. We show the ICM density profiles estimated for SPT-CLJ0304-4921 using a standard X-ray analysis with all 2298 counts available (left) and with a down-sampled event file containing 167 counts in the 0.7-2~keV band (middle). The density profile estimated with a joint X-ray/SZ analysis with 167 counts and the SPT integrated Compton parameter is shown in the right panel. All three profiles are compatible within their uncertainties. However, on the one hand, the relative uncertainty associated with the density profile obtained with a standard X-ray analysis of the down-sampled event file (middle) varies between 95 and 130\% between 10 and 500~kpc. On the other hand, the relative uncertainty associated with the density profile computed with the joint X-ray/SZ analysis (right) varies between 10 and 30\% in the same radius range. On average, we observe that the relative uncertainty on the ICM density profile is decreased by a factor ${\sim}2.5$ in the core and in the outskirts, and by a factor ${\sim}8$ in the intermediate regions around 200~kpc by analyzing jointly the \chandra\ data and the SPT integrated Compton parameter. In the case of SPT-CLJ0304-4921 the mean relative uncertainty between 10 and 500~kpc on the density profile obtained with a standard X-ray analysis of all available counts (left) is 10\%. On average, there is a factor 7 between the number of available counts in the high-S/N and the low-S/N sub-sample (see Appendix~\ref{sec:app_A}). Thus, for a known ICM temperature, we expect a factor $\sqrt{7} \simeq 2.6$ increase of the relative uncertainty on the density profile between the standard X-ray analysis based on all available counts and the joint X-ray/SZ analysis based on the down-sampled event files. As this is indeed the typical factor observed in our analyses, we conclude that the final uncertainties on the ICM density profiles derived from the joint X-ray/SZ analysis are limited by the Poisson fluctuations in the X-ray surface brightness profile. 

\subsection{SZ systematic effects}\label{subsec:sys_eff}

The gain in constraining power demonstrated in \textsection \ref{subsec:gain} comes from the use of the SPT integrated Compton parameter $Y_{SZ}^{0.75'}$ as a constraint of the ICM pressure content. It is therefore essential to ensure that any systematic effect associated with the measurement of this quantity is well characterized.\\
\indent In particular, the high redshift and low mass end of the progenitor sample might be affected by Eddington bias which induces an over-estimation of $Y_{SZ}^{0.75'}$. The corresponding clusters are all part of the SPTpol 100d catalog \citep{hua19}. In Appendix~\ref{sec:app_C}, we compare the integrated Compton estimates for clusters detected in both the SPT-SZ and SPTpol 100d surveys, to estimate the fraction of clusters significantly affected by Eddington bias in the SPTpol 100d sample. It shows that only two clusters out of the 17 SPTpol 100d clusters considered in this work have a $Y_{SZ}^{0.75'}$ estimate lying below the conservative limit below which we consider the SPTpol 100d clusters to be significantly affected by Eddington bias. Furthermore, these two values of $Y_{SZ}^{0.75'}$ are consistent with the considered limit of $3.45\times 10^{-5}~\mathrm{arcmin}^2$. Therefore, Eddington bias is not significantly over-estimating the SPT integrated Compton parameters considered in this work.\\ 
\indent In addition, the SPT integrated Compton parameters are estimated by using the SZ detection centers while our analysis considers the X-ray centroid as a deprojection center. This difference might also over-estimate the values of $Y_{SZ}^{0.75'}$ compared to the ones that would be otherwise obtained by using the X-ray centroids. We estimated the angular distance between the SZ and X-ray centroids for each cluster in the progenitor sample. We find a median deviation of 19~arcsec with a standard deviation of 9~arcsec. As the SPT beam is well characterized by a Gaussian with a FWHM of 1~arcmin, this median difference induces an over-estimation of 12.5\% of $Y_{SZ}^{0.75'}$. As the X-ray surface brightness profile scales as the square of ICM density and the square root of ICM temperature (see Eq.~\ref{eq:XSB_def}), this bias on $Y_{SZ}^{0.75'}$ induces a bias on the ICM density of the order of 1\%. This is negligible given the uncertainties obtained with the joint X-ray/SZ analysis described in \textsection \ref{subsubsec:xsz_method}.\\
\indent Therefore, we consider that the $Y_{SZ}^{0.75'}$ estimates considered in this work are not driving a significant bias on the ICM density profiles obtained from the joint analysis of \chandra\ and SPT data.


\section{Redshift evolution of the ICM core density}\label{sec:results}

After ensuring the validity of the joint X-ray/SZ analysis (see \textsection \ref{sec:performance}), we apply this procedure to all 67 clusters in our sample. We use the down-sampled event files for the high-S/N sub-sample and the original event files for the low-S/N sub-sample to avoid any S/N-driven bias in the final results (see \textsection \ref{subsec:rescale}).\\
\indent At the end of this analysis, we obtain the ICM density profiles of all 67 clusters by applying the same analysis procedure based on event files sharing similar S/N levels. These profiles are shown in the left panel of Fig.~\ref{fig:ne_profs}. We convert each electron density profile into a gas density profile $\rho_g = m_p n_e A/Z$ where $A = 1.397$ and $Z = 1.199$ are the average nuclear mass and charge obtained for a plasma with a metal abundance of $0.3Z_{\odot}$. The gas density profiles are further scaled by the critical density of the universe $\rho_c$ at each cluster redshift. We observe a large scatter in the distribution of almost two orders of magnitude in the core of the clusters while all profiles are compatible within uncertainties at intermediate radii around $0.5$R$_{500}$.\\
\indent We compute the mean gas density profiles in two redshift bins at $z < 0.5$ and $z > 0.5$ in order to maximize the effect induced by redshift evolution while considering similar intervals in lookback time. The two profiles are shown in the right panel of Fig.~\ref{fig:ne_profs} in blue and purple, respectively. We highlight the range of scaled radius $r/\mathrm{R}_{500}$ where $\mathrm{S/N} > 2$ (see \textsection \ref{subsec:rescale}) using two vertical bars. The error bars associated with each profile are given by $\Delta \rho = \sigma \rho / \sqrt{N}$, where $\sigma \rho$ is the scatter of the distribution of gas density profiles in each bin and $N$ is the number of profiles. The mean scaled gas density profiles are fully compatible in the intermediate region at $r/\mathrm{R}_{500} > 0.5$ which shows that the bulk of the gas content evolves self-similarly in this progenitor sample. However, the profiles deviate from each other by about two standard deviations at lower radii. This behavior is consistent with the one observed in Fig.~2 of \cite{mcd17} using a mass-selected sample of clusters. As it is shown in \cite{mcd17}, these features can be explained if clusters are characterized by a core with a non-evolving gas density associated with a self-similarly evolving non-cool core profile. The drop in scaled gas density in the cluster core is thus fully explained by the increased value of the critical density of the universe in the high-redshift bin with respect to the one observed at low redshift. Here, we show that even if we focus on a progenitor-selected sample, cluster cores seem to have had fixed gas densities although their bulk gas content has been growing by a factor ${\sim}4$ within the past 9~Gyr (see \textsection \ref{sec:spt_prog_samp}).\\
\begin{figure*}[t]
\centering
\includegraphics[height=6.3cm]{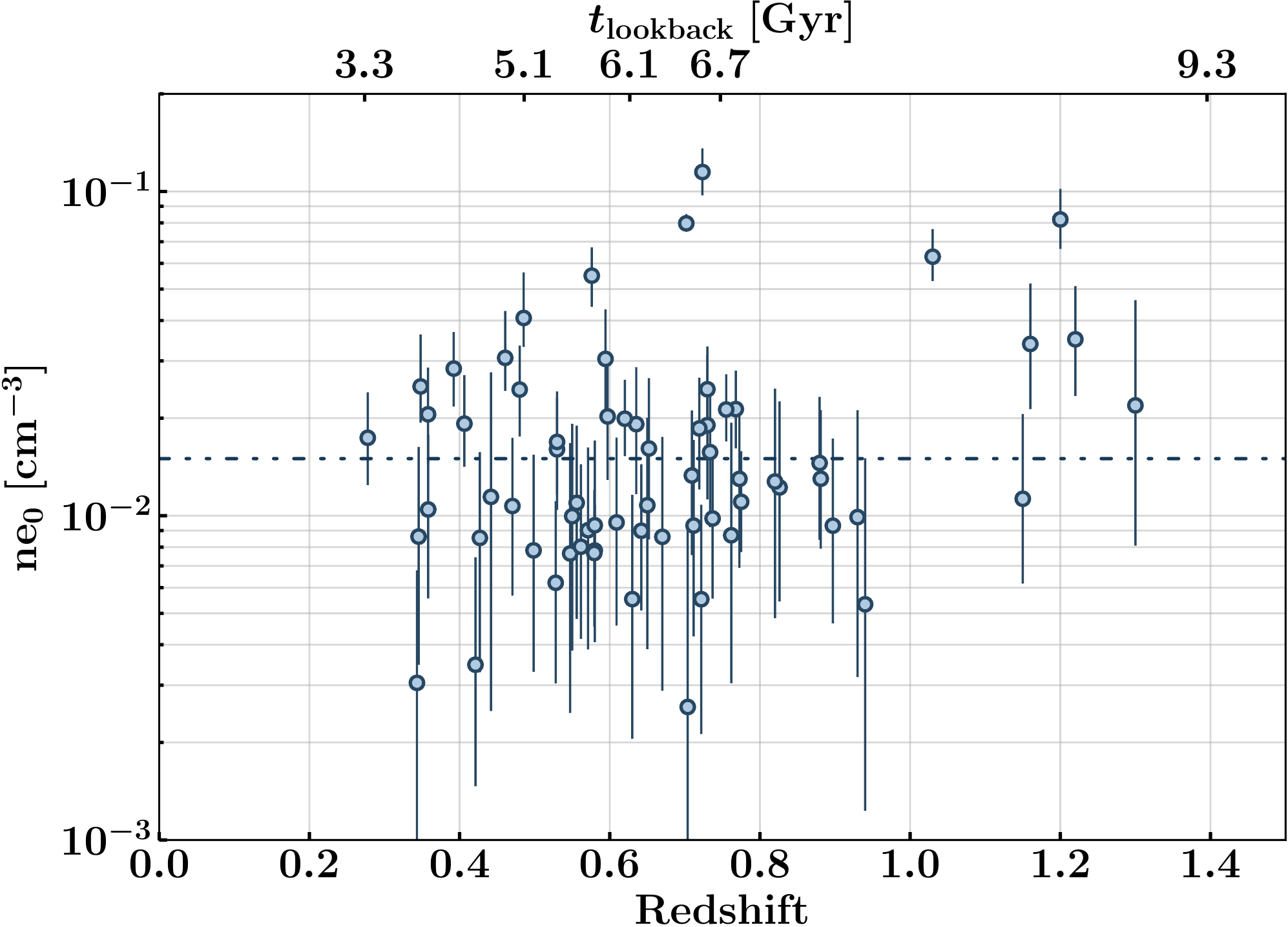}
\hspace{0.5cm}
\includegraphics[height=6.3cm]{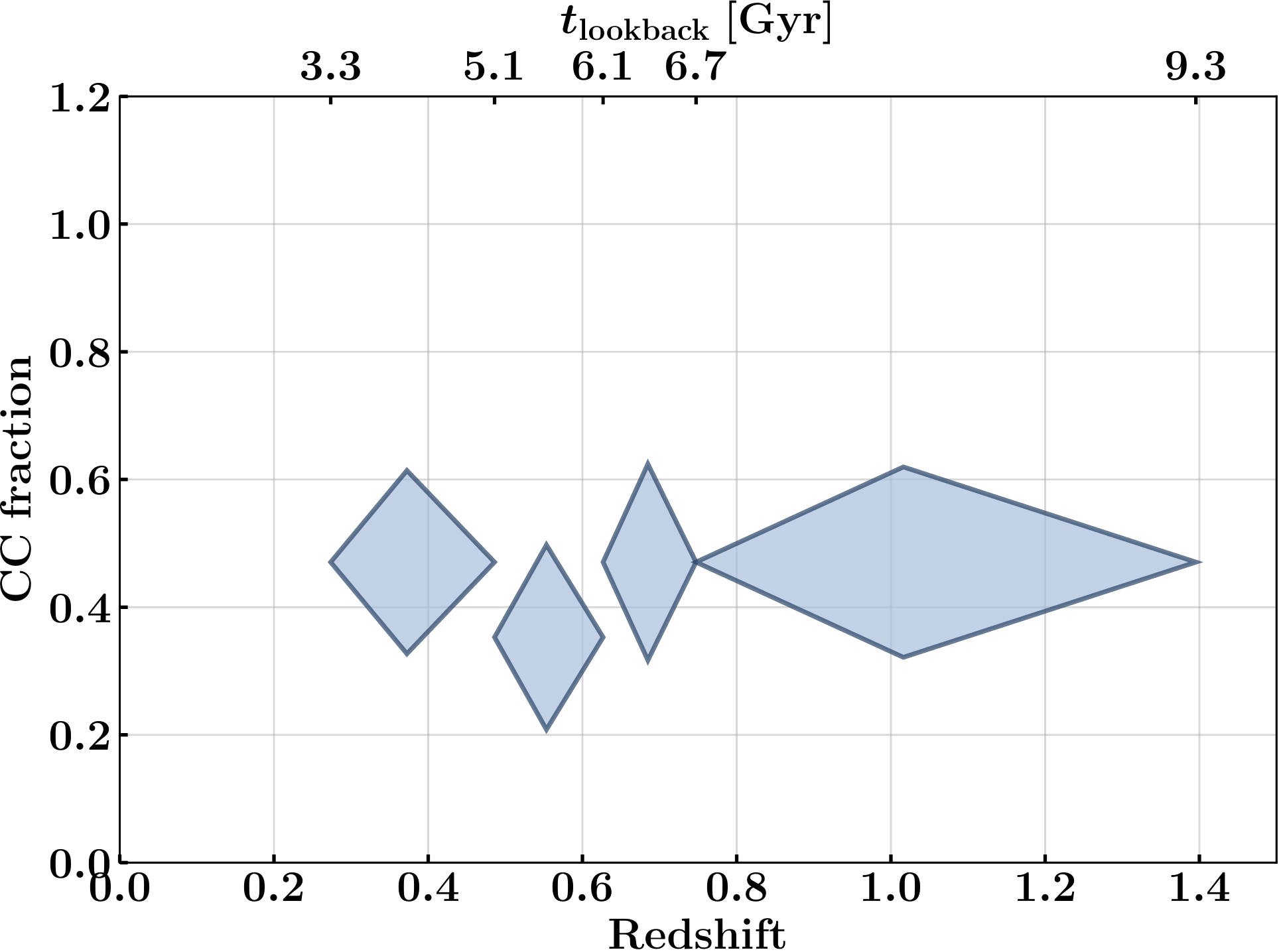}
\caption{{\footnotesize \textbf{Left:} Core electron densities estimated at 10~kpc for each cluster in the progenitor sample as a function of redshift. The dash-dotted line shows the considered boundary between cool core and non-cool core clusters. \textbf{Right:} Cool core fractions estimated by using the core density distribution shown in the left panel in four redshift bins with similar numbers of clusters. The size of the diamonds in each direction gives the $1\sigma$ error bar that also take the binomial uncertainty into account. In both panels, the upper axis gives the lookback times associated with the four redshift bins considered to estimate the cool core fraction (see \textsection \ref{sec:results}.)}}
\label{fig:fcc_z}
\end{figure*}
Based on the previous result, we assume that cluster cores have a fixed size in the following. We estimate the core density of each cluster in the progenitor sample and their associated uncertainties by extracting the value found at 10~kpc in each ICM electron density profile estimated from the joint X-ray/SZ analysis. We show the distribution of ICM core densities as a function of redshift in the left panel of Fig.~\ref{fig:fcc_z} .\\
\indent We realize a profile likelihood ratio analysis \citep[see \emph{e.g.}][]{rup14} in order to test the significance of a linear evolution of the ICM core density. The profile likelihood ratio test statistic allows us to quantify the gap between a constant evolution hypothesis (\emph{i.e.} non-evolving), $H_0$, and a linearly evolving core density hypothesis, $H_1$. It is defined as:
\begin{equation}
q_0 =  -2\mathrm{ln}\left[\frac{\mathscr{L}(n_{e,0},H_0)}{\mathscr{L}(n_{e,0},H_1)}\right],
\end{equation}
where $\mathscr{L}$ is the likelihood of the measured core densities $n_{e,0}$ given the evolution models $H_0$ and $H_1$. It is defined as:
\begin{equation}
-2\mathrm{ln}\mathscr{L} = \sum_{i=1}^{N} \mathrm{ln}(\sigma^2 + \Delta n_{e,0_i}^2) +  \sum_{i=1}^{N} \frac{(n_{e,0_i} - \hat{n}_{e,0}(z_i))^2}{\sigma^2 + \Delta n_{e,0_i}^2},
\end{equation}
where $\Delta n_{e,0}$ is the uncertainty associated with each point, the model $\hat{n}_{e,0}(z) = 10^{az+b}$ is characterized by $a=0$ under $H_0$ and $a \neq 0$ under $H_1$, $b$ is free in both cases, and $\sigma$ is the intrinsic scatter associated with the distribution presented in Fig.~\ref{fig:fcc_z}. We sample the values of the core densities within their uncertainties and realize 10,000 Monte Carlo (MC) realizations of the fit of the distribution with the model $\hat{n}_{e,0}(z)$ under the two hypotheses. For each realization, we compute the significance of the test $Z = \sqrt{q_0}$, following Wilk's theorem. We find that the significance of the test is lower than $3$ for 92.5\% of the MC realizations. The $H_0$ hypothesis is not rejected and the distribution of core densities shown in Fig.~\ref{fig:fcc_z} is thus fully compatible with a non-evolving distribution. We note however, that this analysis does not exclude non-linear models. There is a hint of an increased fraction of high-density cores at $z > 1$ with respect to the distribution observed at $z < 1$. However, there are only 6 clusters at $z > 1$ in the progenitor sample. Therefore, this effect is not significant given the large binomial uncertainty associated with this small sub-sample of 6 clusters.\\
\indent We compute the cool core fraction based on the core ICM densities estimated in four redshift bins in order to study its redshift evolution. The bins are defined between $0.3 < z < 1.3$ in order to contain the same number of clusters, \emph{i.e.} 16 or 17 clusters. We assume that cool core clusters are characterized by a core ICM density $n_{e,0} > 1.5 \times 10^{-2}~\mathrm{cm}^{-3}$, following the results from \cite{hud10}. We estimate the uncertainties associated with the cool core fractions by propagating the measurement errors associated with each data point in the left panel of Fig.~\ref{fig:fcc_z} and by summing the corresponding uncertainty on the cool core fraction in quadrature with the binomial uncertainty derived from \cite{cam13}. The cool core fractions estimated in the four redshift bins along with their corresponding $1\sigma$ uncertainties are shown in the right panel of Fig.~\ref{fig:fcc_z}.\\
\indent We observe that the cool core fraction is not evolving with redshift in the progenitor sample given the size of the bins considered in this work. Although clusters have been growing in mass by a factor four in the past 9~Gyr, there does not appear to be any impact of mass accretion on their gas density content and on the subsequent cool core fraction.


\section{Discussion and perspectives}\label{sec:discussion}

The results obtained from the joint X-ray/SZ analysis described in \textsection \ref{subsubsec:xsz_method} demonstrate that it is not essential to measure ${\sim}1000$ cluster counts in order to estimate the ICM density profile of SZ-selected clusters with relative uncertainties of the order of $20$\%. The integrated Compton parameter is a quantity that is available in most SZ cluster catalogs. It can directly be used as a constraint on the ICM temperature in the fitting procedure of the X-ray surface brightness profile without requiring dedicated analyses of millimeter data. This result opens the possibility to study hundreds of low redshift clusters ($z < 0.5$) with exposures of the order of $1$~ks per cluster. Moreover, together with the increasing sensitivity of SZ cluster surveys \citep{ben14,deb16}, joint X-ray/SZ analyses offer a new path towards the characterization of low mass systems at low redshifts at relatively low cost.\\
\indent The results described in \textsection \ref{sec:results} are consistent with the ones established in previous studies \citep[\emph{e.g.}][]{mcd17,san17} focusing on mass-selected samples containing clusters that follow different evolutionary tracks in the mass-redshift plane (see Fig.~\ref{fig:m_z_plane}). This indicates that cool cores are formed early, at $z > 1.3$, in the structure formation history and stay on average unaffected by AGN feedback during cluster growth. This also implies that cool core disruption by mergers \citep[\emph{e.g.}][]{gom02,dou18,cha20} has to be compensated by cool core restoration mechanisms in timescales that are shorter than the Hubble time \citep{ros11} in order to maintain a constant fraction of cool core clusters with redshift.\\
\indent Few simulations have been used in order to estimate the redshift evolution of the cool core fraction in a progenitor-selected sample of clusters. In \cite{bar18}, the redshift evolution of the cool core fraction measured from the core densities of the IllustrisTNG massive clusters shows a clear positive slope of the order of $0.55\pm 0.10$ with redshift. This is significantly steeper than the evolution found in this work, which is consistent with a null slope. This tends to show that core disruption events in high-redshift cool core clusters occur more frequently or that cool cores are formed much earlier than what is observed in recent simulations. Achieving a better agreement between simulations and observations results on the evolution of the cool core fraction will require improving the underlying galaxy formation model in simulations to take into account the multi-scale mechanisms driving cluster core dynamics.\\
\indent The hint for an increased cool core fraction at $z > 1$ will need to be confirmed by increasing the number of clusters at high redshift between our selection cuts. To this end, we have submitted a cycle 22 \chandra\ proposal, based on the latest version of the SPTpol 100d catalog, in order to add 7 new clusters in our sample at $z > 0.9$. The proposal has been accepted and the observations will be realized in 2021. This will allow us to split the fourth redshift bin in Fig.~\ref{fig:fcc_z} into two parts and test whether the cool core fraction at $z > 1$ significantly deviates from the constant value observed at $z < 1$. If this is the case, a careful treatment of the SPT selection function will need to be realized as gas-poor systems ($f_{\mathrm{gas}} \ll 0.125$) are expected to be more frequently observed at low mass \citep[\emph{e.g.}][]{vik06,koe07,pla13}. If such clusters are found below the SPT detection limit, the population of SPT clusters at $z>1$ might be biased towards cool core systems.\\
\indent The SPT progenitor sample provides a legacy-class resource for the whole cluster science community in the form of a multi-wavelength sample of 67 clusters spanning $0.3 < z < 1.3$ that lie along a common evolutionary track and are the progenitors of well-studied nearby clusters (see \textsection \ref{sec:spt_prog_samp}). Beyond the results presented in this work, the unique properties of this sample will provide many opportunities for additional follow-up studies from new facilities such as the James Webb Space Telescope \citep{gar06}, ALMA \citep{woo09}, the Rubin Observatory \citep{lss09}, EELT \citep{nei18} and SKA \citep{huy13}, ensuring that the legacy of this program would endure for years to come. 


\section{Summary and Conclusions}\label{sec:conclu}

The \chandra\ follow-up of the first generation SPT cluster catalog has yielded tremendous scientific returns. Continued follow-up of the second generation catalogs can now expand cluster science into a new, high-$z$, lower-mass regime. In this paper, we have presented results from a joint X-ray/SZ analysis of 67 clusters selected to be the progenitors of well-known systems such as Perseus and Abell 2390 at $0.3 < z < 1.3$ in the SPT-SZ and SPTpol 100d catalogs. This study allowed us to track the evolution of the ICM core properties over ${\sim}9$~Gyr of cluster growth. We focused our work on the ICM electron density distribution and defer the study of the other ICM thermodynamic properties to a future paper. We summarize the main results of our work below.
\begin{itemize}
\item[$\bullet$] We find that, in this SZ-selected sample of 67 clusters, the number of systems with a spatially flat ICM-induced X-ray emission in the core is too large to consider the X-ray peak as a stable deprojection center for the whole sample. However, the X-ray centroid location is stable with S/N variations and is a more relevant deprojection center for clusters with a disturbed core.\\
\item[$\bullet$] We conduct a joint X-ray/SZ analysis of the \chandra\ surface brightness profile and the SPT integrated Compton parameter in order to push the investigation of the ICM to low mass and high redshift. We show that this procedure allows us to accurately estimate the ICM density profile of all 67 clusters with a relative uncertainty of the order of 20\% without using X-ray spectroscopy. This represents an improvement of a factor ${\sim}5$ with respect to the relative uncertainties obtained with a standard X-ray analysis of the low-S/N clusters.
\item[$\bullet$] Consistent with earlier works, we find that the gas density profile is well modeled by an early-formed ($z > 1.3$) core whose properties remain fixed with redshift in combination with a self-similarly evolving bulk gas distribution. This suggests that mechanical feedback from AGN is occurring in a gentle way in cluster cool cores.
\item[$\bullet$] We find that the redshift evolution of the ICM core density is consistent with a constant in the measured redshift range although clusters have grown in mass by a factor ${\sim}4$ in the past ${\sim}9$~Gyr. We further show that the cool core fraction estimated from the ICM core densities remains constant with redshift. Hydrodynamical simulations tend to prefer an increasing fraction of cool core systems with redshift which is in tension with the results from this work. This suggests that cool cores must have formed earlier or that core disruption mechanisms must occur more frequently at high redshift than what is observed in current simulations.
\end{itemize}
This work highlights that a multi-wavelength approach provides a unique opportunity to uncover the evolution of the gas content within clusters across cosmic history and alleviate inherent degeneracies between ICM properties. There will be at least another decade before the next generation of X-ray observatories such as \emph{Athena} and \emph{Lynx} comes into play. In the time being, we can already push the investigation of the ICM properties to lower mass and higher redshift using the existing X-ray facilities by alleviating the need for precise X-ray spectroscopic measurements to perform a standard X-ray analysis. This work demonstrates that a joint analysis of X-ray and SZ data can accomplish such an endeavor in a sample of clusters with limited resources. The progenitor sample defined in this work has a huge potential that we still need to exploit given its large multi-wavelength coverage in radio, mm, sub-mm, optical/IR, and X-ray. These future studies will pave the way for the next generation of observatories that will trace the properties of the ICM back to the formation of galaxy clusters at $z{\sim}3$.


\section*{Acknowledgements}

\small{F.R. acknowledges financial supports provided by NASA through SAO Award Number SV2-82023 issued by the Chandra X-Ray Observatory Center, which is operated by the Smithsonian Astrophysical Observatory for and on behalf of NASA under contract NAS8-03060. This work was performed in the context of the South-Pole Telescope scientific program. SPT is supported by the National Science Foundation through grants PLR-1248097 and OPP-1852617. Partial support is also provided by the NSF Physics Frontier Center grant PHY-0114422 to the Kavli Institute of Cosmological Physics at the University of Chicago, the Kavli Foundation and the Gordon and Betty Moore Foundation grant GBMF 947 to the University of Chicago. Argonne National Laboratory's work was supported by the U.S. Department of Energy, Office of High Energy Physics, under contract DE-AC02-06CH11357. B.B. is supported by the Fermi Research Alliance LLC under contract no. De-AC02-07CH11359 with the U.S. Department of Energy.}

\appendix

\section{Properties of each cluster in the progenitor sample}\label{sec:app_A}

Below we provide the main properties for the full sample of 67 clusters at $0.3 < z < 1.3$. We divide the sample into two high-S/N and low S/N sub-samples. We describe how the ICM core densities of each cluster is obtained in \textsection \ref{subsec:xsz_analysis}.

\begin{table*}
\begin{tabular}{cccccccc}
\hline
\hline
Name & RA & Dec & $z$ & $M_{500}$ & $n_{e,0}$ & OBSIDs & $\mathrm{N_{counts}}$ \\
 & $[{}^{\circ}]$ & $[{}^{\circ}]$ & & $[10^{14}~\mathrm{M_{\odot}}]$ & $[\mathrm{cm^{-3}}]$ &  & \\
\hline
SPT-CLJ0235-5121 & 38.9390 & -51.3570 & 0.28 & $6.41^{+1.08}_{-1.08}$ & $0.017^{+0.007}_{-0.005}$ & 12262 & 3155 \\
SPT-CLJ0217-5245 & 34.3000 & -52.7515 & 0.34 & $4.42^{+0.89}_{-0.89}$ & $0.003^{+0.004}_{-0.002}$ & 12269 & 1572 \\
SPT-CLJ0555-6406 & 88.8693 & -64.1059 & 0.35 & $7.69^{+1.22}_{-1.22}$ & $0.009^{+0.008}_{-0.005}$ & 13404 & 1387 \\
SPT-CLJ0106-5943 & 16.6163 & -59.7208 & 0.35 & $6.23^{+1.05}_{-1.05}$ & $0.025^{+0.011}_{-0.006}$ & 13468 & 1278 \\
SPT-CLJ2325-4111 & 351.3015 & -41.1964 & 0.36 & $7.55^{+1.20}_{-1.20}$ & $0.021^{+0.008}_{-0.006}$ & 13405 & 1208 \\
SPT-CLJ0348-4515 & 57.0702 & -45.2485 & 0.36 & $6.17^{+1.03}_{-1.03}$ & $0.010^{+0.007}_{-0.005}$ & 13465 & 704 \\
SPT-CLJ0304-4921 & 46.0690 & -49.3574 & 0.39 & $7.57^{+1.20}_{-1.20}$ & $0.028^{+0.008}_{-0.007}$ & 12265 & 2298 \\
SPT-CLJ0013-4906 & 3.3285 & -49.1158 & 0.41 & $7.08^{+1.15}_{-1.15}$ & $0.019^{+0.008}_{-0.005}$ & 13462 & 1411 \\
SPT-CLJ0252-4824 & 43.1991 & -48.4134 & 0.42 & $4.79^{+0.93}_{-0.93}$ & $0.003^{+0.004}_{-0.002}$ & 13494 & 1136 \\
SPT-CLJ2135-5726 & 323.9110 & -57.4393 & 0.43 & $6.15^{+1.02}_{-1.02}$ & $0.009^{+0.007}_{-0.005}$ & 13463 & 1025 \\
SPT-CLJ0330-5228 & 52.4707 & -52.5794 & 0.44 & $6.67^{+1.08}_{-1.08}$ & $0.011^{+0.016}_{-0.009}$ & 893 & 6487 \\
SPT-CLJ0509-5342 & 77.3407 & -53.7024 & 0.46 & $5.06^{+0.89}_{-0.89}$ & $0.031^{+0.012}_{-0.006}$ & 9432 & 1714 \\
SPT-CLJ0655-5234 & 103.9721 & -52.5690 & 0.47 & $5.10^{+0.93}_{-0.93}$ & $0.011^{+0.007}_{-0.005}$ & 13486 & 402 \\
SPT-CLJ2233-5339 & 338.3195 & -53.6530 & 0.48 & $5.48^{+0.98}_{-0.98}$ & $0.024^{+0.009}_{-0.007}$ & 13504 & 1067 \\
SPT-CLJ0334-4659 & 53.5486 & -46.9964 & 0.49 & $5.52^{+0.95}_{-0.95}$ & $0.041^{+0.016}_{-0.008}$ & 13470 & 1392 \\
SPT-CLJ0200-4852 & 30.1405 & -48.8722 & 0.50 & $4.76^{+0.90}_{-0.90}$ & $0.008^{+0.008}_{-0.005}$ & 13487 & 700 \\
SPT-CLJ2035-5251 & 308.7922 & -52.8548 & 0.53 & $6.21^{+1.04}_{-1.04}$ & $0.006^{+0.005}_{-0.003}$ & 13466 & 623 \\
SPT-CLJ0346-5439 & 56.7301 & -54.6481 & 0.53 & $5.47^{+0.94}_{-0.94}$ & $0.016^{+0.007}_{-0.006}$ & 12270 & 728 \\
SPT-CLJ2306-6505 & 346.7241 & -65.0904 & 0.53 & $5.73^{+0.98}_{-0.98}$ & $0.017^{+0.007}_{-0.005}$ & 13503 & 868 \\
SPT-CLJ2335-4544 & 353.7879 & -45.7393 & 0.55 & $6.17^{+1.02}_{-1.02}$ & $0.008^{+0.009}_{-0.005}$ & 13496 & 954 \\
SPT-CLJ0307-5042 & 46.9599 & -50.7044 & 0.55 & $5.26^{+0.93}_{-0.93}$ & $0.010^{+0.009}_{-0.006}$ & 13476 & 1172 \\
SPT-CLJ0232-5257 & 38.1992 & -52.9544 & 0.56 & $5.36^{+0.94}_{-0.94}$ & $0.011^{+0.008}_{-0.006}$ & 12263 & 765 \\
SPT-CLJ0456-5116 & 74.1191 & -51.2791 & 0.56 & $5.09^{+0.89}_{-0.89}$ & $0.008^{+0.006}_{-0.004}$ & 13474 & 1308 \\
SPT-CLJ2148-6116 & 327.1771 & -61.2803 & 0.57 & $4.46^{+0.83}_{-0.83}$ & $0.009^{+0.007}_{-0.005}$ & 13488 & 819 \\
SPT-CLJ2331-5051 & 352.9634 & -50.8642 & 0.58 & $5.60^{+0.92}_{-0.92}$ & $0.055^{+0.012}_{-0.011}$ & 18241 & 3087 \\
SPT-CLJ0307-6225 & 46.8321 & -62.4301 & 0.58 & $5.06^{+0.90}_{-0.90}$ & $0.008^{+0.004}_{-0.003}$ & 12191 & 681 \\
SPT-CLJ0256-5617 & 44.1017 & -56.2976 & 0.58 & $4.54^{+0.85}_{-0.85}$ & $0.009^{+0.008}_{-0.005}$ & 14448 & 808 \\
SPT-CLJ2245-6206 & 341.2568 & -62.1186 & 0.58 & $5.40^{+0.94}_{-0.94}$ & $0.008^{+0.003}_{-0.001}$ & 13499 & 1154 \\
SPT-CLJ2232-5959 & 338.1433 & -59.9986 & 0.59 & $5.55^{+0.97}_{-0.97}$ & $0.030^{+0.013}_{-0.009}$ & 13502 & 1093 \\
SPT-CLJ0033-6326 & 8.4700 & -63.4443 & 0.60 & $4.72^{+0.88}_{-0.88}$ & $0.020^{+0.010}_{-0.007}$ & 13483 & 476 \\
SPT-CLJ0559-5249 & 89.9334 & -52.8244 & 0.61 & $5.78^{+0.95}_{-0.95}$ & $0.010^{+0.008}_{-0.005}$ & 13117 & 1198 \\
SPT-CLJ0123-4821 & 20.7973 & -48.3564 & 0.62 & $4.46^{+0.87}_{-0.87}$ & $0.020^{+0.006}_{-0.005}$ & 13491 & 1215 \\
SPT-CLJ0426-5455 & 66.5179 & -54.9187 & 0.63 & $5.17^{+0.90}_{-0.90}$ & $0.006^{+0.006}_{-0.003}$ & 13472 & 578 \\
SPT-CLJ0243-5930 & 40.8625 & -59.5193 & 0.64 & $4.58^{+0.85}_{-0.85}$ & $0.019^{+0.010}_{-0.008}$ & 13484 & 1039 \\
SPT-CLJ0542-4100 & 85.7093 & -41.0021 & 0.64 & $5.16^{+0.94}_{-0.94}$ & $0.009^{+0.005}_{-0.004}$ & 914 & 1407 \\
SPT-CLJ2218-4519 & 334.7445 & -45.3176 & 0.65 & $5.31^{+0.92}_{-0.92}$ & $0.011^{+0.009}_{-0.007}$ & 13501 & 791 \\
SPT-CLJ2222-4834 & 335.7109 & -48.5784 & 0.65 & $5.42^{+0.93}_{-0.93}$ & $0.016^{+0.010}_{-0.008}$ & 13497 & 873 \\
SPT-CLJ0352-5647 & 58.2398 & -56.7990 & 0.67 & $4.24^{+0.81}_{-0.81}$ & $0.009^{+0.009}_{-0.006}$ & 13490 & 481 \\
SPT-CLJ0000-5748 & 0.2502 & -57.8099 & 0.70 & $4.72^{+0.50}_{-0.59}$ & $0.079^{+0.004}_{-0.004}$ & 18238,18239,19695 & 4731 \\
SPT-CLJ0310-4647 & 47.6348 & -46.7869 & 0.71 & $4.31^{+0.83}_{-0.83}$ & $0.013^{+0.008}_{-0.006}$ & 13492 & 570 \\
SPT-CLJ0102-4603 & 15.6713 & -46.0676 & 0.72 & $4.49^{+0.85}_{-0.85}$ & $0.006^{+0.005}_{-0.003}$ & 13485 & 675 \\
SPT-CLJ2043-5035 & 310.8220 & -50.5929 & 0.72 & $4.53^{+0.86}_{-0.86}$ & $0.115^{+0.021}_{-0.017}$ & 18240 & 3392 \\
SPT-CLJ0324-6236 & 51.0516 & -62.5986 & 0.73 & $4.97^{+0.86}_{-0.86}$ & $0.025^{+0.009}_{-0.007}$ & 13137 & 500 \\
SPT-CLJ2301-4023 & 345.4709 & -40.3887 & 0.73 & $4.81^{+0.86}_{-0.86}$ & $0.019^{+0.012}_{-0.008}$ & 13505 & 837 \\
SPT-CLJ2352-4657 & 358.0677 & -46.9576 & 0.73 & $4.42^{+0.83}_{-0.83}$ & $0.016^{+0.009}_{-0.006}$ & 13506 & 1083 \\
SPT-CLJ0406-4805 & 61.7287 & -48.0831 & 0.74 & $4.61^{+0.83}_{-0.83}$ & $0.010^{+0.007}_{-0.004}$ & 13477 & 430 \\
SPT-CLJ0528-5300 & 82.0210 & -52.9964 & 0.77 & $3.65^{+0.73}_{-0.73}$ & $0.021^{+0.007}_{-0.005}$ & 11874,11747,12092 & 687 \\
SPT-CLJ2359-5009 & 359.9309 & -50.1689 & 0.77 & $3.60^{+0.71}_{-0.71}$ & $0.011^{+0.005}_{-0.003}$ & 9334,11742,11864,11997 & 695 \\
SPT-CLJ0058-6145 & 14.5809 & -61.7695 & 0.83 & $4.36^{+0.81}_{-0.81}$ & $0.012^{+0.010}_{-0.007}$ & 13479 & 563 \\
SPT-CLJ0533-5005 & 83.4061 & -50.0959 & 0.88 & $3.79^{+0.73}_{-0.73}$ & $0.013^{+0.008}_{-0.005}$ & 12002,12001,11748 & 886 \\
\hline
\end{tabular}
\caption{{\footnotesize Properties of the 50 clusters in the high-S/N sub-sample along with information about the corresponding \chandra\ observations. The average number of counts in the 0.7-2~keV band is 1282.}}
\label{tab:high_sn}
\end{table*}

\begin{table*}
\begin{tabular}{cccccccc}
\hline
\hline
Name & RA & Dec & $z$ & $M_{500}$ & $n_{e,0}$ & OBSIDs & $\mathrm{N_{counts}}$ \\
 & $[{}^{\circ}]$ & $[{}^{\circ}]$ & & $[10^{14}~\mathrm{M_{\odot}}]$ & $[\mathrm{cm^{-3}}]$ &  & \\
\hline
SPT-CLJ2355-5156 & 358.8441 & -51.9508 & 0.70 & $3.24^{+0.39}_{-0.47}$ & $0.003^{+0.005}_{-0.003}$ & 19760 & 128 \\
SPT-CLJ2355-5258 & 358.9327 & -52.9777 & 0.71 & $2.87^{+0.40}_{-0.46}$ & $0.009^{+0.009}_{-0.005}$ & 19757 & 105 \\
SPT-CLJ2329-5831 & 352.4730 & -58.5293 & 0.72 & $3.87^{+0.44}_{-0.52}$ & $0.018^{+0.008}_{-0.007}$ & 19762 & 266 \\
SPT-CLJ2320-5233 & 350.1235 & -52.5633 & 0.76 & $2.68^{+0.38}_{-0.44}$ & $0.023^{+0.007}_{-0.005}$ & 22953,21552 & 191 \\
SPT-CLJ0000-6020 & 0.0344 & -60.3382 & 0.76 & $2.90^{+0.39}_{-0.46}$ & $0.009^{+0.008}_{-0.005}$ & 19758 & 110 \\
SPT-CLJ2328-5533 & 352.1807 & -55.5670 & 0.77 & $3.08^{+0.39}_{-0.46}$ & $0.012^{+0.008}_{-0.005}$ & 19759 & 167 \\
SPT-CLJ0001-5440 & 0.4091 & -54.6719 & 0.82 & $3.37^{+0.40}_{-0.48}$ & $0.012^{+0.010}_{-0.007}$ & 19761 & 107 \\
SPT-CLJ2343-5024 & 355.8371 & -50.3993 & 0.88 & $3.29^{+0.39}_{-0.47}$ & $0.014^{+0.009}_{-0.007}$ & 19764 & 146 \\
SPT-CLJ2304-5718 & 346.1079 & -57.3069 & 0.90 & $2.53^{+0.38}_{-0.42}$ & $0.008^{+0.006}_{-0.004}$ & 21551 & 171 \\
SPT-CLJ2311-5820 & 347.9923 & -58.3445 & 0.93 & $2.25^{+0.40}_{-0.41}$ & $0.010^{+0.011}_{-0.007}$ & 19763 & 179 \\
SPT-CLJ2325-5116 & 351.3850 & -51.2852 & 0.94 & $2.08^{+0.36}_{-0.43}$ & $0.005^{+0.010}_{-0.005}$ & 19753 & 170 \\
SPT-CLJ2335-5434 & 353.8824 & -54.5867 & 1.03 & $2.11^{+0.37}_{-0.42}$ & $0.060^{+0.014}_{-0.008}$ & 23027,21553,23159 & 365 \\
SPT-CLJ0002-5557 & 0.5144 & -55.9667 & 1.15 & $2.62^{+0.36}_{-0.41}$ & $0.012^{+0.012}_{-0.006}$ & 21550 & 195 \\
SPT-CLJ2259-5301 & 344.8229 & -53.0330 & 1.16 & $1.98^{+0.34}_{-0.41}$ & $0.040^{+0.017}_{-0.012}$ & 21556,22022,22046,22047, & 224 \\
 & & & & & & 22048,22053,22054 & \\
SPT-CLJ2334-5308 & 353.5153 & -53.1406 & 1.20 & $2.25^{+0.38}_{-0.38}$ & $0.081^{+0.019}_{-0.015}$ & 19756,23026,23181,21705 & 208 \\
SPT-CLJ2336-5252 & 354.0805 & -52.8727 & 1.22 & $2.45^{+0.35}_{-0.41}$ & $0.033^{+0.015}_{-0.013}$ & 19886,21706,23127 & 229 \\
SPT-CLJ2323-5752 & 350.8817 & -57.8798 & 1.30 & $1.92^{+0.35}_{-0.40}$ & $0.021^{+0.033}_{-0.015}$ & 21555,23051,23052, & 164 \\
 & & & & & & 23121,23136 & \\
\hline
\end{tabular}
\caption{{\footnotesize Properties of the 17 clusters in the low-S/N sub-sample along with information about the corresponding \chandra\ observations. The average number of counts in the 0.7-2~keV band is 184.}}
\label{tab:low_sn}
\end{table*}

\section{Typical X-ray spectrum with 180 counts}\label{sec:app_B}

Below, we show the X-ray spectrum extracted in a single annulus mapping the radius range $0.15 \mathrm{R}_{500} < r <  \mathrm{R}_{500}$ using the down-sampled event file associated with SPT-CLJ0304-4921. This event file is characterized by a total number of counts of 180 in the 0.7-2~keV band in a circular region of radius R$_{500}$ centered on the X-ray centroid. The spectrum is background dominated at energies $E > 1.5$~keV. Furthermore, it is fully compatible with a background-only spectrum between 0.7 and 2~keV. The temperature of the ICM is thus compatible with 0~keV.

\begin{figure*}[h!]
\centering
\includegraphics[height=7cm]{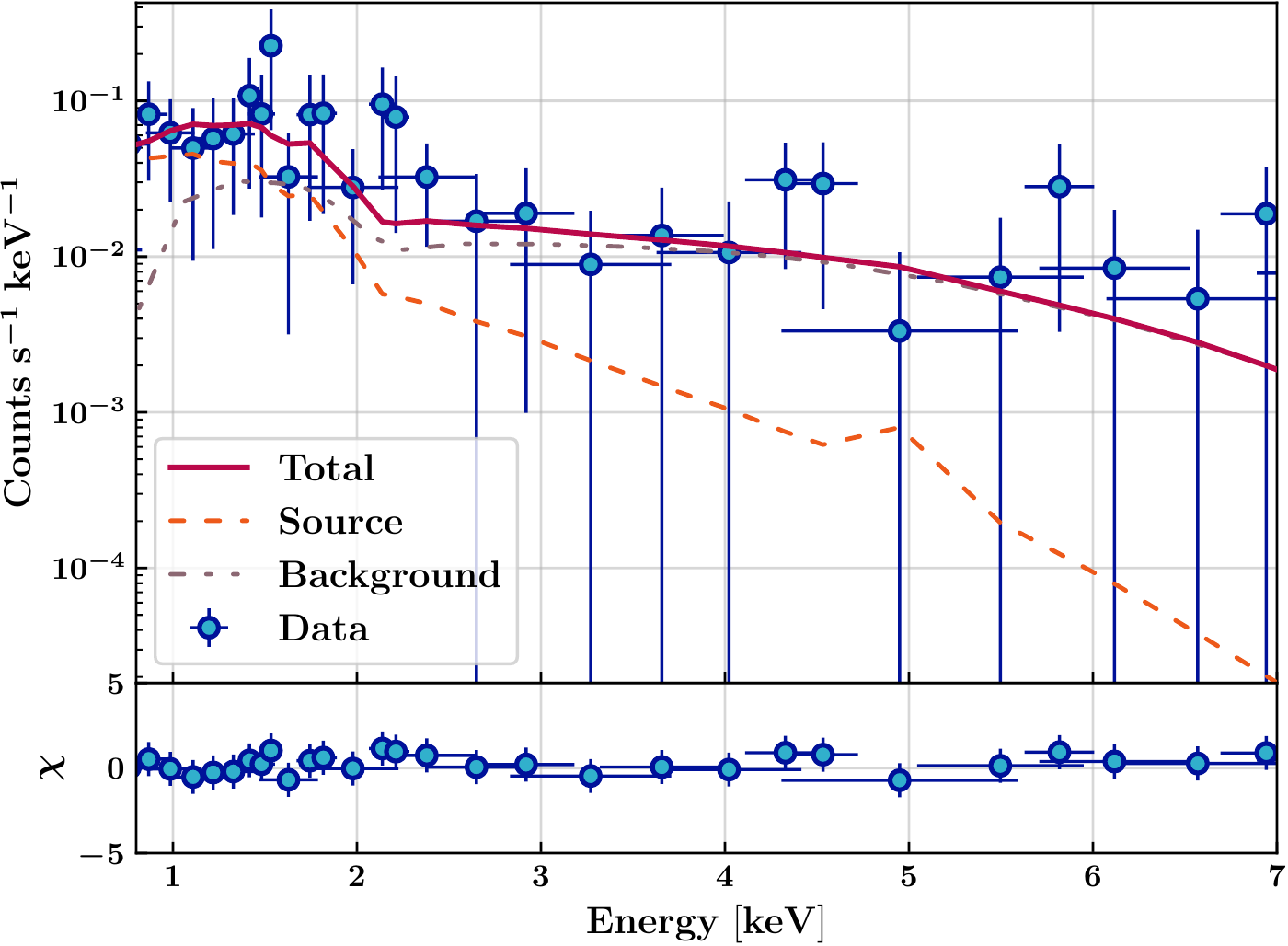}
\caption{{\footnotesize X-ray spectrum of SPT-CLJ0304-4921 extracted from a down-sampled event file (see \textsection \ref{subsec:rescale}). The best-fit model (see \textsection \ref{subsec:standard_X}) is shown in red. It is the combination of a background model (grey) and an ICM emission model (orange). The weighted difference between the data and the model is shown in the lower panel. The significance of these residuals is also lower than $3\sigma$ in all energy bins if we consider the bakground model only as a best-fit model.}}
\label{fig:spec_180}
\end{figure*}

\section{Impact of Eddington bias}\label{sec:app_C}

\begin{figure*}[h!]
\centering
\includegraphics[height=9.5cm]{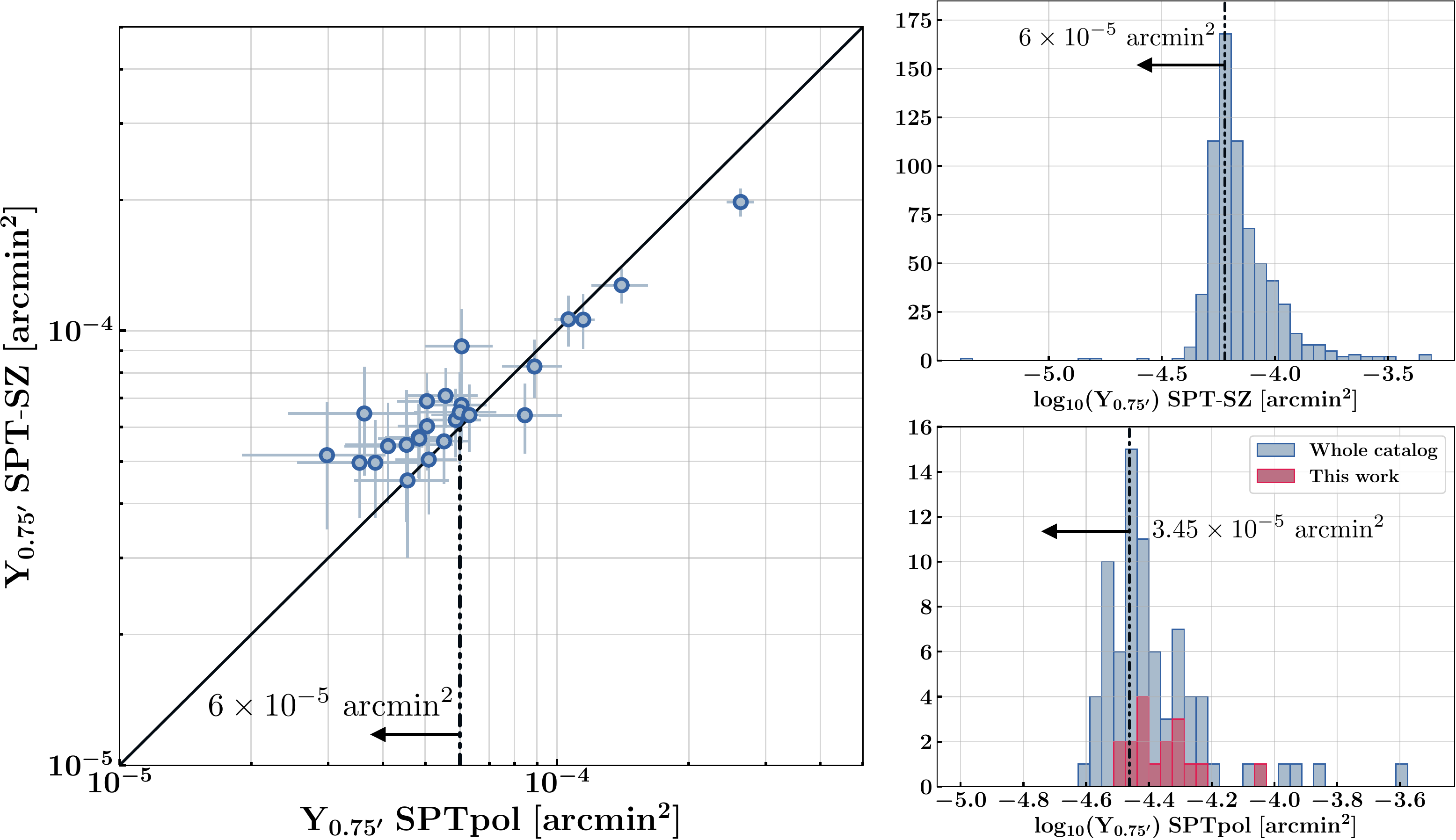}
\caption{{\footnotesize \textbf{Left:} Comparison between the values of $Y_{0.75'}$ measured for 25 clusters that are detected both in the SPT-SZ \citep{ble15} and SPTpol 100d \citep{hua19} surveys. \textbf{Right:} Histograms showing the values of $Y_{0.75'}$ measured for all clusters in the SPT-SZ (top) and SPTpol 100d (down) catalogs. We also show the $Y_{0.75'}$ values of the 17 SPTpol clusters considered in this work (red). In all panels, the limit below which we consider that Eddington bias is significant is shown with a vertical dash-dotted line. }}
\label{fig:edd_bias}
\end{figure*}

We estimated the impact of Eddington bias on the values of the integrated Compton parameter $Y_{0.75'}$ of the SPTpol clusters by assuming that the fractions of clusters that are affected by this bias in the SPT-SZ and SPTpol catalogs are similar. The assumption is valid as the signal-to-noise thresholds considered for cluster detection are almost identical in both catalogs ($\xi = 4.5$ in \cite{ble15} and $\xi=4.6$ in \cite{hua19}). In Fig.~\ref{fig:edd_bias}, we show a comparison between the $Y_{0.75'}$ estimates for the 25 clusters detected in both the SPT-SZ and SPTpol 100d surveys. Based on SZ field scaling factors \citep[see \emph{e.g.}][]{deh16}, the SPTpol 100d field is 1.9 times deeper than the same field in the SPT-SZ survey considering cluster detection ability. We consider a conservative estimate of $6\times 10^{-5}~\mathrm{arcmin}^2$ to define a lower limit from which clusters in the SPT-SZ catalog are not significantly affected by Eddington bias. We define this threshold as the value from which the $Y_{0.75'}$ estimates in both the SPT-SZ and SPTpol 100d catalogs are consistent with the equality line (black solid line in Fig.~\ref{fig:edd_bias}). In the upper right panel of Fig.~\ref{fig:edd_bias}, we show the distribution of all SPT-SZ $Y_{0.75'}$ values. We find that 35\% of clusters in this catalog have an integrated Compton parameter that is below $6\times 10^{-5}~\mathrm{arcmin}^2$. In the lower right panel of Fig.~\ref{fig:edd_bias}, we show the distribution of the $Y_{0.75'}$ values obtained for all SPTpol 100d clusters (blue) along with those associated with the 17 SPTpol 100d clusters considered in this work (red). If we assume that 35\% of the SPTpol catalog is significantly affected by Eddington bias, we obtain a conservative limit of $3.45\times 10^{-5}~\mathrm{arcmin}^2$ below which we consider that the $Y_{0.75'}$ estimates are over-estimated. Only two clusters considered in this work satisfy this condition and both of them have a $Y_{0.75'}$ value that is compatible with this limit.

\end{document}